\documentclass[journal]{IEEEtran}
\ifCLASSINFOpdf
   \usepackage[pdftex]{graphicx}
 \else
   \usepackage[dvips]{graphicx}
\fi
\usepackage{soul}
\usepackage{amsmath}
\usepackage{amsfonts}

\usepackage{amssymb}
\usepackage{pifont}
\newcommand{\cmark}{\ding{51}}
\newcommand{\xmark}{\ding{55}}

\usepackage{multirow}
\usepackage{makecell}
\usepackage{colortbl}
\usepackage{xcolor}
\usepackage{array}
\usepackage{verbatim}
\usepackage{booktabs}
\usepackage{orcidlink}
\ifCLASSOPTIONcompsoc
  \usepackage[caption=false,font=normalsize,labelfont=sf,textfont=sf]{subfig}
\else
  \usepackage[caption=false,font=footnotesize]{subfig}
\fi
\usepackage{xfrac}

\usepackage{url}

\DeclareUnicodeCharacter{0301}{\'{e}}

\begin{document}
\title{Speaker anonymization using orthogonal Householder neural network}

\author{Xiaoxiao Miao \orcidlink{0000-0002-6645-6524}, \IEEEmembership{Member, IEEE},
        Xin Wang \orcidlink{0000-0001-8246-0606}, \IEEEmembership{Member, IEEE},
        Erica Cooper \orcidlink{0000-0002-2978-2793}, \IEEEmembership{Member, IEEE}, 
       Junichi Yamagishi \orcidlink{0000-0003-2752-3955}, \IEEEmembership{Senior Member, IEEE},
        Natalia Tomashenko \orcidlink{0000-0002-7125-2382},~\IEEEmembership{Member,~IEEE}
\thanks{This study is supported by JST CREST Grants (JPMJCR18A6 and JPMJCR20D3), MEXT KAKENHI Grants (21K17775, 21H04906, 21K11951, 22K21319), and the VoicePersonal project (ANR-18-JSTS-0001). 

The associate editor coordinating the review of this manuscript and approving it for publication was Prof. Zhizheng Wu.
\textit{(Corresponding author: Xiaoxiao Miao.)}}
\thanks{Xiaoxiao Miao, Xin Wang, Erica Cooper and Junichi Yamagishi are with the National Institute of Informatics, 2-1-2 Hitotsubashi Chiyoda-ku, Tokyo 101- 8340, Japan (e-mail: {xiaoxiaomiao, wangxin, ecooper, jyamagis}@nii.ac.jp).}
\thanks{Natalia Tomashenko is with Laboratoire Informatique d’Avignon (LIA), Avignon University, France (email: natalia.tomashenko@univ-avignon.fr)}}

\markboth{Journal of \LaTeX\ Class Files,~Vol.~14, No.~8, August~2015}%
{Shell \MakeLowercase{\textit{et al.}}: Bare Demo of IEEEtran.cls for IEEE Journals}

\maketitle

\begin{abstract}

Speaker anonymization aims to conceal a speaker's identity while preserving content information in speech.
Current mainstream neural-network speaker anonymization systems disentangle speech into prosody-related, content, and speaker representations.
The speaker representation is then anonymized by a selection-based speaker anonymizer that uses a mean vector over a set of randomly selected speaker vectors from an external pool of English speakers.
However, the resulting anonymized vectors are subject to severe privacy leakage against powerful attackers, reduction in speaker diversity, and language mismatch problems for unseen-language speaker anonymization.
To generate diverse, language-neutral speaker vectors, this paper proposes an anonymizer based on an orthogonal Householder neural network (OHNN).
Specifically, the OHNN acts like a rotation to transform the original speaker vectors into anonymized speaker vectors, which are constrained to follow the distribution over the original speaker vector space.
 A basic classification loss is introduced to ensure that anonymized speaker vectors from different speakers have unique speaker identities.
To further protect speaker identities, an improved classification loss and similarity loss are used to push original-anonymized sample pairs away from each other.
Experiments on VoicePrivacy Challenge datasets in English and the \textit{AISHELL-3} dataset in Mandarin demonstrate the proposed anonymizer's effectiveness.

\end{abstract}


\section{Introduction}
\label{sec:intro}
\IEEEPARstart{S}{peech} technology enables machines to recognize, analyze, and understand human speech, which facilitates human-machine communication and offers great convenience in our daily lives. 
Despite its prominent advantages, it suffers from voice privacy leakage, which allows for intrusion upon or tampering with a speaker’s private information.
For instance, by using advanced speaker \cite{reynolds1995speaker,miao2021d}, dialect \cite{ali2019mgb, miao2021variance}, pathological condition \cite{dibazar2002feature, schuller2013interspeech}, or other types of speech attribute recognition systems, attributes such as a speaker's identity, geographical origin, and health status can easily be captured from speech recordings.
Moreover, advanced speech synthesis techniques enable resynthesis, cloning, or conversion of a speaker’s identity information to access personal voice-controlled devices \cite{vestman2020voice,jin2008voice,Das2020}. 
In this paper, we are especially interested in speaker anonymization, 
which is a user-centric voice privacy solution to conceal a speaker's identity without degrading intelligibility and naturalness \cite{fang2019speaker, tomashenko2020introducing,tomashenko2021voiceprivacy}.
This task was standardized by the VoicePrivacy Challenge (VPC) committee \cite{tomashenko2020introducing,tomashenko2021voiceprivacy,tomashenko2022voiceprivacy}, which held challenges in 2020 and 2022, to advance the development of voice privacy preservation techniques.

Several approaches to protect speaker privacy are based on digital signal processing (DSP) methods \cite{tomashenko2020introducing,tomashenko2021voiceprivacy, patino2020speaker, gupta2020design,dubagunta2020adjustable,tavi2022improving,vpc22tsm}, which modify instantaneous speech characteristics such as the pitch, spectral envelope, and time scaling. 
State-of-the-art anonymization approaches have borrowed ideas from neural speech conversion and synthesis, mainly focusing on disentangled latent representation learning \cite{fang2019speaker,jin2009voice, qian2018hidebehind, huang2021defending, magarinos2017reversible,srivastava2022privacy,meyer2022anonymizing,yao22_spsc} via two hypotheses.
The first is that speech can be explicitly decomposed into content, speaker identity, and prosodic (intonation, stress, and rhythm) representations. Here, the speaker identity is a statistical time-invariant representation throughout an utterance, whereas content and prosodic information vary over time.
The second hypothesis is that a speaker's identity representation carries most of his or her private information. 
Thus, generated speech using original content, prosodic, and anonymized speaker representations can suppress the original identity information (privacy) while maintaining intelligibility and naturalness (utility).

A general framework for disentanglement-based speaker anonymization involves the following components.

\textit{Fine-grained disentangled representation extraction from original speech:}
Here, extraction entails three aspects:
(i) Content feature extraction. Low-dimensional phonetic bottleneck features are typically extracted from an intermediate layer of a language-specific automatic speech recognition neural acoustic model (ASR AM) \cite{povey2018semi, peddinti2015time}. 
This type of content encoder is trained in a supervised manner using transcribed English training data. As the objective is to obtain accurate linguistic representations, the effectiveness is severely limited when applied to a different language. Content encoders based on self-supervised learning (SSL) can overcome this limitation thanks to being trained in a self-supervised manner using unlabeled training data. Specifically, they can provide general content representations not dependent on the language, thus enabling robust anonymization of speech data even for unseen languages.
(ii) Prosody-related feature extraction to obtain the fundamental frequency, i.e., F0.
(iii) Speaker embedding extraction. A speaker vector is extracted either from an automatic speaker verification (ASV) system based on a time-delay neural network (TDNN)  \cite{snyder2018x}, or from a more effective ASV system based on emphasized channel attention, propagation and aggregation in TDNN (ECAPA-TDNN) \cite{desplanques2020ecapa}.
 
\textit{Speaker representation anonymization}:
The core idea of a speaker vector anonymizer is to hide original speaker information while preserving the diversity among different speakers.
A widely used anonymizer is based on the selection and averaging of speaker vectors \cite{Srivastava2020DesignCF,srivastava2022privacy}. 
Given a large set of speaker vectors, the anonymizer finds the $N$ farthest candidate vectors away from an input original vector. It then randomly selects $N^* < N$ vectors among the $N$ farthest ones and utilizes their average as a pseudo-speaker vector to replace the original speaker vector. 
The large set of speaker vectors, called an external pool, has to be loaded by the anonymizer during anonymization.

\textit{Anonymized speech synthesis}: 
An anonymized speaker vector with the original fundamental frequency and content features is passed to a speech waveform generation model to synthesize high-quality anonymized speech. The speech synthesis model can be a traditional text-to-speech pipeline model---a speech synthesis acoustic model (SS AM) and a neural source filter- (NSF-) based vocoder \cite{wang2019neural}---or a unified HiFi-GAN \cite{kong2020hifi}.

Despite confirmation of this approach's effectiveness \cite{tomashenko2020introducing,tomashenko2021voiceprivacy,miao22_odyssey}, 
there remains much room for improvement for different attack scenarios and unseen language anonymization.
Previous works \cite{tomashenko2020introducing,tomashenko2021voiceprivacy,miao22_odyssey,srivastava2022privacy} have suggested that the most significant performance bottleneck for the current mainstream approach is the selection-based speaker anonymizer, whose performance significantly depends on the distribution of the external pool and how pseudo-speakers are selected from the pool. (i) For English speaker anonymization \cite{tomashenko2020introducing,tomashenko2021voiceprivacy,tomashenko2022voiceprivacy}, the performance of speaker verifiability has gradually decreased against more powerful attackers. Additionally, voice distinctiveness is significantly degraded by anonymization.
(ii) For unseen-language (e.g., Mandarin) speaker anonymization, pseudo-speaker representations are generated from an external English speaker vector pool, and the resulting language mismatch increases the character error rate (CER) \cite{miao22_odyssey,miao2022analyzing}.

\begin{figure}[t]
\centering
\includegraphics[width=0.8\columnwidth]{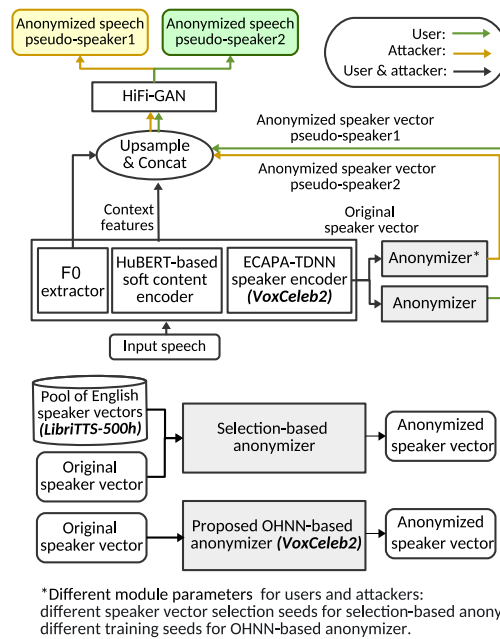}
\caption{Architecture of an SSL-based SAS, with selection- and OHNN-based anonymizers.}
\label{fig:structure-overall}
\end{figure}

Following this pipeline of disentanglement-based anonymization, with special consideration of the selection-based approach's problems, we propose a novel speaker anonymization system (SAS) based on an orthogonal Householder neural network (OHNN). As shown in the lower part of Fig. \ref{fig:structure-overall},  the OHNN-based anonymizer generates distinctive anonymized speaker vectors that can protect privacy under all attack scenarios and can successfully be adapted to unseen-language speaker anonymization without severe language mismatch.
Specifically, original speaker vectors are rotated to anonymized ones by an OHNN, which is a linear transformation with orthogonality.
This module ensures that the anonymized speaker vectors follow the distribution over the original speaker vector space.
To discourage overlap between anonymized speakers and other speakers, we use a classification loss based on an additive angular margin softmax (AAM) and cross-entropy to train the OHNN, and we assign different target class labels to the original and anonymized speaker vectors of different speakers.  
This encourages the anonymized vectors to not overlap with any other speakers, regardless of whether they are original or anonymized.
To further push original-anonymized sample pairs away from each other, an improved classification loss called weighted AAM (w-AAM) and a cosine similarity loss are used.

The main contributions of this work are as follows:
\begin{itemize}
\item We propose an OHNN-based anonymizer that transforms original speaker vectors into anonymized ones with carefully designed training constraints.
We show empirically that these anonymized speaker vectors are diverse and language-neutral.
\item We visualize the cosine similarities between pairs of speaker vectors extracted from the generated speech of users and different attackers. These generated speech are obtained using the commonly used selection-based anonymizer and our OHNN-based anonymizer.
The results show that our proposed method effectively reduces the privacy leakage against different attackers and improves the diversity of anonymized speakers.
We conducted experiments on VPC English datasets and the \textit{AISHELL-3} Mandarin datasets. Our findings show that the proposed model can be successfully adapted to both a matched language condition (i.e., English) and a mismatched language condition where the target language (Mandarin) is not included in the training database. The proposed anonymizer achieved a competitive performance under all attack scenarios in terms of privacy and utility metrics. Under the \textit{Semi-informed} condition, our proposed methods achieved better results for English speaker anonymization than all the submissions to VPC2022 \cite{chen22_spsc, gaznepoglu22_spsc, meyer2022anonymizing, mawalim22_spsc, khamsehashari22_spsc, yao22_spsc}.
\end{itemize}

\begin{figure*} 
    \centering
     \includegraphics[width=0.7\linewidth]{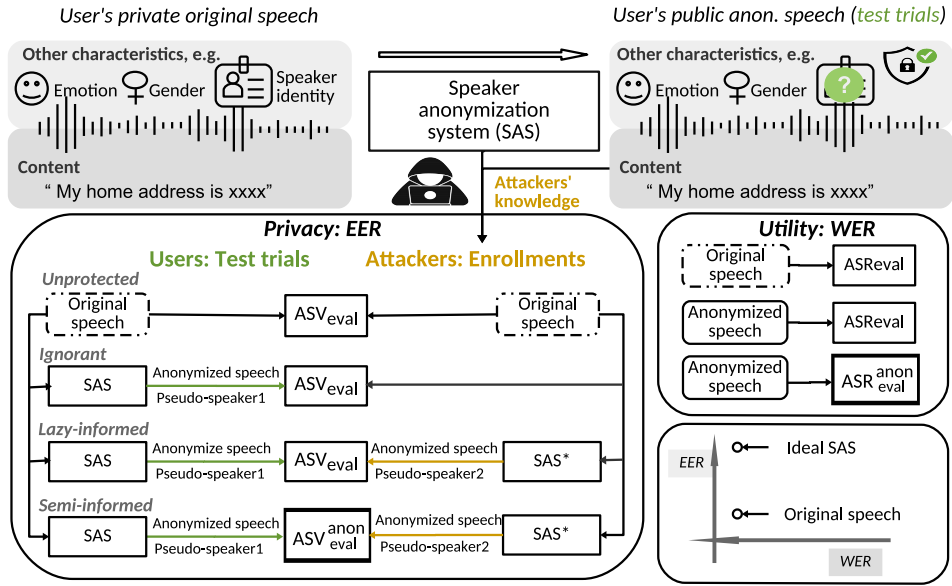}
     \caption{Speaker anonymization task. A user anonymizes original speech to hide his or her identity before publication, and attackers use biometric (ASV) technology and knowledge of the anonymization method to re-identify the original speaker's identity. }
     \label{fig:eval-types}
\end{figure*}

\section{Related Work}
\label{sec:related work}
In this section, we introduce the VPC's official design, which provides the setting for this study, including definitions of specific goals, attack models, and objective evaluation metrics. 
We also overview existing speaker anonymization approaches and their limitations.

\subsection{The VoicePrivacy Challenges}
The VPC formulates the speaker anonymization task as a game between users and attackers, as shown in Fig. \ref{fig:eval-types}.
A user publishes anonymized data, called \textit{test trials}, after applying an SAS to his or her original private speech.
According to the VPC evaluation plan \cite{tomashenko2022voiceprivacy}, an SAS should:
\begin{itemize}
    \item output an anonymized speech waveform;
    \item conceal the speaker's identity from different attackers;
    \item keep content and other paralinguistic attributes unchanged to maintain intelligibility and naturalness;
    \item ensure all test trials from the same speaker are attributed to the same pseudo-speaker, while test trials from different speakers have different pseudo-speakers \footnote{This is called speaker-level anonymization. A different approach known as utterance-level anonymization assigns different pseudo-identities to different utterances of the same original speaker. In this work, we follow the VPC protocol and utilize speaker-level anonymization.}
 \end{itemize}

\subsubsection{Attack Models and Objective Evaluation Metrics}
\label{sec:evaluation}
 
\paragraph{Privacy metric}
To assess the ability to protect a speaker's identity in different scenarios, the ASV performance in terms of the equal error rate (EER) is computed as the primary privacy metric by using language-matched ASV evaluation models.
This metric is calculated under the four attack models shown in the lower left of Fig. \ref{fig:eval-types}. 
The attackers are assumed to have access to a few original or anonymized utterances for each speaker, called \textit{enrollment} utterances, and to have different levels of knowledge about the SAS:

\begin{itemize}
\item \textit{Unprotected}: No anonymization is applied, and attackers verify the original test trials against the original enrollment data by using an ASV system trained on the original dataset, denoted {$ASV_\text{eval}$}.
\item \textit{Ignorant}: Attackers are unaware of the anonymization strategy used for the test trial utterances; instead, they use the original enrollment data and {$ASV_\text{eval}$} to infer a speaker's identity.
\item \textit{Lazy-informed}: Attackers use a similar SAS without accurate parameters to anonymize their enrollment data, and they use {$ASV_\text{eval}$}  to detect a speaker's identity.
\item \textit{Semi-informed}: The only difference from \textit{Lazy-informed} is that the attackers use {$ASV_\text{eval}^\text{anon}$}, a more powerful version trained on anonymized speech, to reduce the mismatch between the original and anonymized speech and infer the speaker's identity.
\end{itemize} 

\paragraph{Primary utility metric} 
To assess how well speech content is preserved in anonymized speech, the ASR performance in terms of the word error rate (WER) is computed as a primary utility metric by using language-matched ASR evaluation models.
As illustrated in the lower right of Fig.~\ref{fig:eval-types}, two ASR models are trained in the same way to decode the anonymized data: $ASR_\text{eval}$, trained on the original data, and $ASR_\text{eval}^\text{anon}$, trained on the anonymized data.
This enables exploration of whether speech content can be maintained better by simply retraining with similarly anonymized data.

\paragraph{Secondary utility metric}
To assess and visualize the preservation of voice distinctiveness, the gain of voice distinctiveness metric, $\text G_\text{VD}$ \cite{noe2020speech,noe2021csl}, is computed.
Precisely, $M=( M(i,j))_{1 \le i \le N,1 \le j \le N}$ is a voice similarity matrix for $N$ speakers, where the similarity value $ M(i,j)$ for speakers $i$ and $j$ is formulated as follows:

\begin{equation}
\small
     M(i,j) = \mathrm{sigmoid}\left({\frac{1}{n_{i}n_{j} }
    \displaystyle\sum_{\substack{1 \le k \le n_{i} \text{ and } 1 \le l \le n_{j} \\ k\neq l \text{ if } i=j } 
    }{\text{LLR}(x^{(i)}_{k},x^{(j)}_{l})}}\right),
    \label{equ:: M}
\end{equation}

Here, $n_{i}$ and $n_j$ are the numbers of utterances for each speaker; and $\text{LLR}(x^{(i)}_{k},x^{(j)}_{l})$ is the log-likelihood ratio obtained by comparing the $k$-th utterance of the $i$-th speaker with the $l$-th utterance of the $j$-th speaker. These LLR scores are computed by probabilistic linear discriminant analysis (PLDA) \cite{ioffe2006probabilistic} of the $ASV_\text{eval}$ model trained on the original data.

Three matrices are constructed from the original (o) and anonymized (a) data:
$M_\text{oo}$ from the original data,  $M_\text{oa}$ from the original and anonymized data, and $M_\text{aa}$ from the anonymized data.
The diagonal dominance $D_\text{diag}(M)$ is computed as the absolute difference between the mean values of diagonal and  off-diagonal elements:

\begin{equation}
\small
    D_{\text{diag}}(M)\hspace{-0.7mm}=\hspace{-0.7mm}\displaystyle
    \Bigg|
    \sum_{1\leq i \leq N} \frac{ M(i,i)}{N}
    \displaystyle
    - 
    \sum_{\substack{1 \le j \le N \text{ and } 1 \le k \le N \\j \neq k}}
    \frac{ M(j,k)}{N(N-1)}
    \Bigg|.
    \label{eq:ddiag}
\end{equation}

Next, $G_{\text{VD}}$ \cite{noe2020speech} is defined as the diagonal dominance ratio of the two matrices:
\begin{equation}
 G_{\text{VD}} = 10\log_{10}  \frac{D_\text{diag}(M_\text{aa})}{D_\text{diag}(M_\text{oo})},   
\end{equation}

Here, a gain of $G_{\text{VD}}=0$ dB indicates that voice distinctiveness is preserved on average after anonymization, while a gain above or below 0 dB corresponds respectively to an average increase or decrease in voice distinctiveness.

An ideal anonymization system should achieve high EERs (close to 50\%) in the \textit{Ignorant}, \textit{Lazy-informed}, and \textit{Semi-informed} scenarios to protect the speaker's information. In addition, the WER should be as low as for the original speech, and $G_{\text{VD}}$ should be close to 0 dB to preserve voice distinctiveness.

\subsection{Existing Speaker Anonymization Approaches} \label{sec:sas}

\subsubsection{Digital Signal Processing (DSP) Methods}
\label{sec:dsp}
A simple approach \cite{patino2020speaker} that does not require training data is to change speaker attributes with distortion of the spectral envelope by using McAdams coefficients \cite{mcadams1984spectral} to randomly shift the positions of formant frequencies.  
Widening of formant peaks \cite{gupta2020design} further distorts the spectral envelope.
Data-driven formant modification can also be applied by using the formant statistics of desired speakers \cite{dubagunta2020adjustable} or time-scale algorithms \cite{vpc22tsm}.
Phonetically controllable anonymization \cite{tavi2022improving} modifies a speaker's vocal tract and voice source features, with a focus on F0 trajectories.
Although these methods perceptually manipulate the speech signal, previous works have indicated that powerful attackers can effortlessly recover speaker identities \cite{tomashenko2020introducing, tomashenko2021voiceprivacy, srivastava2020evaluating}.

\subsubsection{Disentangled Representation Methods}
A typical approach based on disentangled representation learning, called x-vector based anonymization, is used as the primary baseline in the VPC \cite{fang2019speaker,tomashenko2020introducing,tomashenko2021voiceprivacy,tomashenko2022voiceprivacy}.
It extracts speaker representations and linguistic features by using a pretrained TDNN-based ASV system \cite{snyder2018x} and ASR AM based on a factorized time-delay neural network (TDNN-F), respectively.
Then, to hide the original speaker's information, a selection-based speaker anonymizer \cite{Srivastava2020DesignCF} replaces the original x-vector with the mean vector of a set of randomly selected speaker vectors from an external pool of English speakers.
Specifically, given a centroid of source speaker vectors from one speaker, the cosine distance is used to find the 200 farthest centroids in an external speaker vector pool, and 100 of those are randomly selected and averaged to obtain an anonymized speaker vector \cite{Srivastava2020DesignCF}.
Finally, an SS AM generates mel-filterbank features from the anonymized pseudo x-vector, F0, and linguistic features, and an NSF-based waveform generator synthesizes anonymized speech. 

Because this disentanglement-based method is more effective at protecting speaker identities than the DSP-based methods discussed in Section \ref{sec:dsp}  \cite{srivastava2020evaluating, tomashenko2021voiceprivacy}, most speaker anonymization studies have followed a similar framework. Improvements mainly come from two sources:

\textit{Improved speech disentanglement}:
Some works \cite{mawalim2022speaker,champion2022disentangled, shamsabadi2022differentially} have argued that the disentangled linguistic information extracted from the language-specific ASR AM and F0 still contain speaker information. Accordingly, they modify the F0 and linguistic information to remove the residual speaker identity. 

\textit{Improved speaker vector anonymization}:
Other researchers have modified the original x-vector in ways that increase the privacy protection ability.
Perero-Codosero \MakeLowercase{\textit{et al.}} \cite{perero2022x} transformed an original x-vector to an anonymized one by using an autoencoder with an adversarial training strategy to suppress speaker, gender, and accent information. This requires labels for the speaker identity, gender, and nationality.
Turner \MakeLowercase{\textit{et al.}} \cite{turner2022generating} sampled anonymized x-vectors from a Gaussian mixture model in a space reduced by principal component analysis (PCA) over an external pool of speakers, which preserves the distributional properties of the original x-vectors.
There have been recent attempts to generate a target pseudo-speaker for speaker anonymization in the systems submitted to the VoicePrivacy Challenge 2022. For example, Meyer \MakeLowercase{\textit{et al.}} \cite{meyer2022anonymizing} utilized a generative adversarial network to generate artificial speaker embeddings, where the anonymization stage requires a manual search to find vectors that are dissimilar to the anonymized one. Yao \MakeLowercase{\textit{et al.}} \cite{yao22_spsc} proposed using a look-up table (LUT)-based method to generate pseudo-speaker embeddings, along with an average of randomly selected speaker embeddings from the real speakers. However, it suffers from limited variability in the anonymized voices. Chen \MakeLowercase{\textit{et al.}} \cite{chen22_spsc} proposed a method for distorting an input speech signal by adding adversarial noise designed to hide the original speaker identity.

Most of the existing approaches are limited in two aspects. First, they use an ASR-based content extractor that requires large amounts of transcribed English training data. Such an ASR-based content extractor is ineffective for speaker anonymization in unseen languages.
Our previous work alleviates this issue by using an SSL-based content extractor \cite{miao22_odyssey}. As shown in Fig. \ref{fig:structure-overall}, this SSL-based SAS consists of a HuBERT-based soft content encoder \cite{van2021comparison}, an ECAPA-TDNN speaker encoder \cite{desplanques2020ecapa}, an F0 extractor, and a HiFi-GAN decoder \cite{kong2020hifi}. It does not require text transcriptions or any other language-specific resources, and it has demonstrated the ability to anonymize speech data with reasonable performance even if the data is in a language not included in the training data.
However, it suffers from a remaining limitation of selection-based anonymizers according to previous results \cite{tomashenko2020introducing,tomashenko2021voiceprivacy,tomashenko2022voiceprivacy,miao22_odyssey,miao2022analyzing}: the distribution of the external speaker pool significantly affects anonymized speakers, and the averaging of vectors from the speaker pool reduces voice distinctiveness. 

\begin{figure}[t]
 \centering
 \includegraphics[width=0.8\columnwidth]{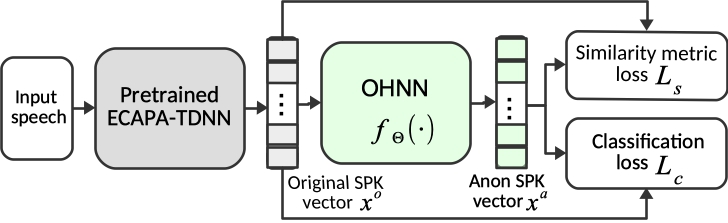}
\caption{ Framework for an OHNN-based anonymizer. 
The $\mathbf{x}^{o}$ are original speaker representations extracted from a pretrained ECAPA-TDNN,
which then pass through a transfer module $f(\cdot)$ to produce the corresponding anonymized speaker representations $\mathbf{x}^{a}$.
The $\mathbf{x}^{o}$ and $\mathbf{x}^{a}$ are trained as different speakers.
After training, $\mathbf{x}^{a}$ is used as a pseudo-speaker vector to synthesize speech.}
\label{fig:reform_nn}
\end{figure}

\section{Proposed OHNN-Based Anonymizer}
\label{sec:nn-based}
To mitigate the problems with existing approaches, we propose the OHNN-based anonymizer shown in Fig. \ref{fig:reform_nn}. 
Hence, this section formulates speaker anonymization as a constrained optimization problem, describes a general form of the proposed anonymizer, and explains the implementation details.  

\subsection{Problem Formulation} 
\label{seq:problem_def}
The training set $\{(\mathbf{x}_i^o, {y}_i^o)\}_{i=1}^{M}$ comprises $M$ speaker vector $\mathbf{x}_i^o$ and the corresponding speaker label ${y}_i^o$. The speaker vector $\mathbf{x}_i^o \in \mathbb{R}^d$ is a $d$-dimensional segment-level speaker embedding obtained from an ECAPA-TDNN pretrained on the original audio waveform. $\mathbf{x}_i^o$ follows an unknown distribution $\mathbf{x}_i^o \sim p_{\mathbf{x}^o}$. 

Anonymized speaker vectors $\mathbf{{x}}_i^a \in \mathbb{R}^d $ are obtained by transforming $\mathbf{x}_i^o$ with a function $f_\Theta: \mathbb{R}^{d}\rightarrow\mathbb{R}^{d}$, written as follows:
\begin{align}
  \mathbf{x}^a =  f_\Theta(\mathbf{x}^o).
\end{align}
Accordingly, the anonymized speaker vectors follow another distribution $\mathbf{x}_i^a \sim p_{\mathbf{x}^a}$ or $\mathbf{x}_i^a \sim p_{f_\Theta(\mathbf{x}^o)}$.

An ideal speaker anonymization method should meet at least three constraints:
\begin{itemize}
\item Speaker privacy protection: $\mathbf{x}_i^o$ and $\mathbf{x}_i^a$ are dissimilar to hide the original speaker identity. More specifically, in the context of VPC, $\mathbf{x}_i^o$ and $\mathbf{x}_i^a$ are dissimilar to the extent that the anonymized speech generated using $\mathbf{x}_i^a$ is recognized as being a different speaker by the attackers' ASV.
\item Speaker diversity: $\mathbf{x}_i^a$ has a unique speaker identity ${{y}}_i^a$ to maintain the diversity of anonymized speech across different speakers.
\item Distribution similarity: $\mathbf{x}_i^a \sim p_{\mathbf{x}^a}$ satisfies the same distribution as $\mathbf{x}_i^o$ to maintain the naturalness of the original speech.
\end{itemize}

The above constraints can be formulated as an optimization problem:

\begin{align}
\begin{split}
\label{eqn:general_objective}
(\Theta,  \Psi)^*  & =   \arg \min_{\Theta, \Psi} \mathbb{E}_{{\{\mathbf{x}^o, y^o\} \in D}} \Big[\lambda \mathcal{L}_s\big(\mathbf{x}^o, f_\Theta(\mathbf{x}^o)\big) \\
 & \quad + \mathcal{L}_c \big(y^o, g_\Psi (\mathbf{x}^o); y^a, g_\Psi(f_\Theta(\mathbf{x}^o))\big) \Big] ,
\end{split}\\
 \label{eqn:general_constraint}
 &\text{s.t. } \mathcal{D}{\big(p_{\mathbf{x}^o},   p_{f_\Theta(\mathbf{x}^o)}\big)} < \epsilon ,
 \end{align}
 where $\lambda$ is a hyperparameter to balance the multi-objective function.
$\mathcal{L}_s$ is a similarity metric to optimize $\Theta$ by minimizing the similarity of the original-anonymized pair, which ideally makes the original and anonymized speech be recognized as different speakers by the attackers' ASV.

Next, $g_\Psi(\cdot)$ denotes the classifier layer, and $\mathcal{L}_c$ is its classification loss function to optimize $\Theta$ and $\Psi$ by minimizing the discrepancy between the sets of desired outputs, $y^o, y^a$, and predicted outputs, $g_\Psi (\mathbf{x}^o), g_\Psi(\mathbf{x}^a)$. The outputs may be defined for a multi-speaker classification task in which the original and corresponding anonymized speaker vectors are intentionally treated as different target speaker classes. This means that all speaker vectors after anonymization are treated as different classes, as well as different classes from the original speakers to maintain speaker diversity.

Finally, $\mathcal{D}{\big(p_{\mathbf{x}^o},   p_{f_\Theta(\mathbf{x}^o)}\big)}$ is the divergence between distributions of $\mathbf{x}$ included in a training database before and after anonymization. This term ensures similarity between the distributions of the anonymized and original speaker vectors, with some tolerance $\epsilon$. The Kullback–Leibler divergence (KLD) or other types of divergence are applicable.

\subsection{General Form of Proposed Anonymizer}
Finding a direct solution of  Eqs.~(\ref{eqn:general_objective}) and (\ref{eqn:general_constraint}) for an arbitrarily designed DNN-based $f_\Theta$ is difficult. 
Here, we propose an anonymizer that, with a few assumptions, always satisfies the constraint in Eq.~(\ref{eqn:general_constraint}) regardless of the value of $\Theta$. In such a case, $\Theta$ and $\Psi$ can be optimized via Eq.~(\ref{eqn:general_objective}) and a conventional gradient descent method.

Let $\boldsymbol{\mu}_{\mathbf{x}^o} \in \mathbb{R}^{d}$ and $\mathbf{\Sigma}_{\mathbf{x}^o} \in \mathbb{R}^{d\times{d}}$ be the mean and covariance matrix of $p_{\mathbf{x}^o}$, respectively. 
Our proposed anonymizer $f_\Theta(\cdot)$ can be written as follows:
\begin{equation}
    \mathbf{x}^{a} = f_\Theta(\mathbf{x}^{o}) = \mathbf{L}_{\mathbf{x}^{o}}^{-1}\mathbf{W}\mathbf{L}_{\mathbf{x}^{o}}(\mathbf{x}^{o} - \boldsymbol{\mu}_{\mathbf{x}^o}) + \boldsymbol{\mu}_{\mathbf{x}^o},
    \label{eqn:general_definition}
\end{equation}
where $\mathbf{L}_{\mathbf{x}^{o}}$ is a whitening matrix\footnote{$\mathbf{L}_{\mathbf{x}^{o}}$ is a whitening matrix. It can be derived from $\mathbf{\Sigma}_{\mathbf{x}^o}$ by a matrix decomposition method used in, e.g., PCA or Cholesky whitening \cite{kessy2018optimal}.} that satisfies $\mathbf{L}_{\mathbf{x}^{o}}^{-1}{\mathbf{L}_{\mathbf{x}^{o}}^{-1}}^{\top} = \mathbf{\Sigma}_{\mathbf{x}^o}$, and $\mathbf{W}\in
\mathbb{R}^{d\times d}$ is an orthogonal matrix that 
satisfies $\mathbf{W}\mathbf{W}^\top = \mathbf{W}^\top\mathbf{W} = \mathbf{I}$.
While $\boldsymbol{\mu}_{\mathbf{x}^{o}}$ and $\mathbf{L}_{\mathbf{x}^{o}}$ are determined by the data distribution, the values of $\mathbf{W}$ are learned via Eq.~( \ref{eqn:general_objective}).

Before introducing the parameterization and optimization of $\mathbf{W}$, we show that the proposed anonymizer satisfies $\mathcal{D}{\big(p_{\mathbf{x}^o},   p_{f_\Theta(\mathbf{x}^o)}\big)} = 0$ given that $p_{\mathbf{x}^o}$ is a Gaussian distribution  $\mathcal{N}(\boldsymbol{\mu}_{\mathbf{x}^o}$, $\mathbf{\Sigma}_{\mathbf{x}^o})$\footnote{Being Gaussian is a desirable but not absolutely required condition to ensure $\mathcal{D}{\big(p_{\mathbf{x}^o},   p_{f_\Theta(\mathbf{x}^o)}\big)} = 0$, but many types of speaker vectors can be assumed to follow a multivariate Gaussian distribution in the high dimensional space. One example is the length-normalized i-vector \cite{garcia2011analysis}. Another example is the ECAPA-TDNN speaker vectors, which can be well modeled using PLDA with Gaussian distributions \cite{wang22r_interspeech}.} We first decompose
Eq.~(\ref{eqn:general_definition}) into three steps:
\begin{itemize}
    \item Centering and whitening: $\tilde{\mathbf{x}}^{o} = \mathbf{L}_{\mathbf{x}^{o}}(\mathbf{x}^{o} - \boldsymbol{\mu}_{\mathbf{x}^o})$,
    \item Rotation:  $\tilde{\mathbf{x}}^{a} = \mathbf{W}\tilde{\mathbf{x}}^{o}$,
    \item De-whitening and de-centering: $\mathbf{x}^{a} = \mathbf{L}_{\mathbf{x}^{o}}^{-1}\tilde{\mathbf{x}}^{a} + \boldsymbol{\mu}_{\mathbf{x}^o}$.
\end{itemize}
The centered and whitened speaker vector  $\tilde{\mathbf{x}}^{o}$ obviously follows a normal distribution $\tilde{\mathbf{x}}^{o} \sim \mathcal{N}(\mathbf{0}, \mathbf{I})$. As $\mathbf{W}$ is an orthogonal matrix, $\tilde{\mathbf{x}}^{a}$ also follows a normal distribution $ \mathcal{N}(\mathbf{W}\mathbf{0}, \mathbf{W}\mathbf{W}^{\top}) = \mathcal{N}(\mathbf{0}, \mathbf{I})$. Through the affine transformation in the last step, we know that $\mathbf{x}^{a} \sim \mathcal{N}(\boldsymbol{\mu}_{\mathbf{x}^o}, \mathbf{L}_{\mathbf{x}^{o}}^{-1} {\mathbf{L}_{\mathbf{x}^{o}}^{-1}}^{\top}) = \mathcal{N}(\boldsymbol{\mu}_{\mathbf{x}^o}, \mathbf{\Sigma}_{\mathbf{x}^{o}})$. Hence, the defined anonymizer does not change the distribution, i.e., $\mathcal{D}\big(p_{\mathbf{x}^o},   p_{f_\Theta(\mathbf{x}^o)}\big) = 0$.

The above explanation also reveals the core idea of our proposed anonymizer: while it does not change the overall distribution, each speaker vector is rotated through an orthogonal transformation. The anonymized $\mathbf{x}^a$ is guaranteed to be different from the original $\mathbf{x}^o$ as long as $\mathbf{W} \neq \mathbf{I}$. While an infinite number of orthogonal matrices can be applied for rotation, the optimal $\mathbf{W}$ with respect to the criterion in Eq.~(\ref{eqn:general_objective}) must be estimated through an optimization process.

In real applications, $\boldsymbol{\mu}_{\mathbf{x}^o}$ and $\mathbf{\Sigma}_{\mathbf{x}^o}$ of the test set data are unknown. They can be estimated by collecting multiple samples from the test domain if it is possible. Otherwise, we can either use the statistics from the training set or make some simplifications. Through preliminary experiments, we found an effective, simplified form:
\begin{equation}
\mathbf{x}^{a} = f_\Theta ( \mathbf{x}^{o}) = \mathbf{W}(\mathbf{x}^{o} - \boldsymbol{\mu}_{\mathbf{x}^o}^{\text{train}}) + \boldsymbol{\mu}_{\mathbf{x}^o}^{\text{train}},
\end{equation}
where $\boldsymbol{\mu}_{\mathbf{x}^o}^{\text{train}}$ is the mean of the speaker vectors in the training set, and $\mathbf{\Sigma}_{\mathbf{x}^o}$ is assumed to be an identity matrix.

\begin{figure}[!t]
\centering
\subfloat[][Random orthogonal Householder (ROH)]{\includegraphics[height=4cm]{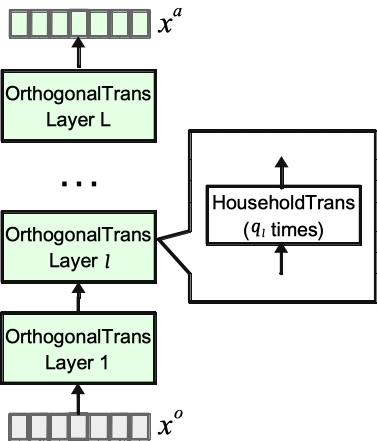}}
\hspace{0.5cm}
\subfloat[][Learnable orthogonal Householder (LOH)]{\includegraphics[height=4cm]{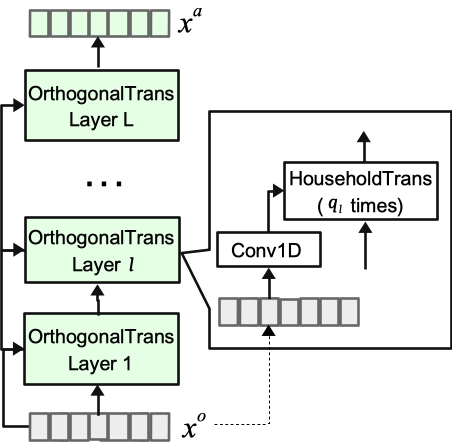}}\vspace{0.0cm}\\
\caption{Two types of OHNN-based anonymizers.}
\label{fig:nn-anon}
\end{figure}

\subsection{Rotation Matrix Using Householder Reflection}
We now need a specific way to parameterize $\mathbf{W}$ to guarantee that the learned $\mathbf{W}$ through gradient descent is orthogonal. 
While many methods can be used, we found that one based on a Householder reflection \cite{householder1958unitary} is efficient for DNNs.
Without loss of generality, assume that $\mathbf{W}$ is a product of multiple orthogonal matrices: 
\begin{align}
\label{eqn:ohnn}
  \mathbf{W} &=\mathbf{W}_1\mathbf{W}_l...\mathbf{W}_L,
\end{align}
where each matrix $\mathbf{W}_l\in\mathbb{R}^{d\times{d}}$ is given by  
\begin{align}
\mathbf{W}_l &= \mathbf{H}_{q_l} \, \mathbf{H}_{q_l-1}\, ... \,\mathbf{H}_1, ~~ q_l \leq d,
 \label{eqn:ohnn_2}
\end{align}
Here, each sub-matrix $\mathbf{H}_{q_l}$ is constructed with a Householder reflection \cite{householder1958unitary} given a non-zero vector $\mathbf{v}_{q_l}\in\mathbb{R}^{d}$ as follows:
\begin{align}
\label{eqn:ohnn_3}
 \mathbf{H}_{q_l}  &= \mathbf{I}  - \frac{2}{\mathbf{v}_{q_l}^\top\mathbf{v}_{q_l}} \mathbf{v}_{q_l}\mathbf{v}_{q_l}^\top.
\end{align}

The resulting $\mathbf{H}$ is known to be an orthogonal matrix for any non-zero vector $\mathbf{v}_{q_l}$, i.e., $\mathbf{H}^\top\mathbf{H}=\mathbf{H}\mathbf{H}^\top=\mathbf{I}$ and $\mathbf{H}\neq \mathbf{I}, \forall \mathbf{v}_{q_l} \neq \mathbf{0}$. Accordingly, $\mathbf{W}_l$ and $\mathbf{W}$ are orthogonal and guaranteed not to be the identity matrix.

Equations (\ref{eqn:ohnn}-\ref{eqn:ohnn_3}) allow us to parameterize $\mathbf{W}$ as $\{\cdots, \mathbf{v}_{q_l}, \cdots\}$. 
We further propose two implementations, which differ in how they compute $\mathbf{v}$:
\begin{enumerate}
    \item[1)] \textit{Random orthogonal Householder (ROH) reflection}:  
    $\mathbf{v}$ is treated as a learnable free parameter, i.e., $\Theta=\{\cdots, \mathbf{v}_{q_l}, \cdots\}$, and each $\mathbf{v}$ is randomly initialized and optimized using Eq.~(\ref{eqn:general_objective}). The anonymization process is illustrated in Fig.~\ref{fig:nn-anon}(a). 
    \item[2)] \textit{Learnable orthogonal Householder (LOH) reflection}: 
    Each $\mathbf{v}$ is transformed from a small NN given the input $\mathbf{x}^o$. In such a case, $\Theta$ is the set of the trainable weights in a set of small NNs. Fig.~\ref{fig:nn-anon}(b) illustrates an implementation in which each DNN has a single 1D convolution layer with 192 output channels and a kernel size of $3 \time 3$. 
\end{enumerate}
While both implementations ensure that the transformation matrix $\mathbf{W}$ is orthogonal, the first approach assumes a global transformation for all the input speaker vectors. In contrast, the latter approach assumes that the transformation matrix varies according to the input.

\subsection {Loss Functions}

Before delving into the details of the loss functions, we describe how to build batch data for an OHNN-based anonymizer.
Let $N$ be the batch size and $C$ be the number of original speakers.
Each mini-batch comprises $N/2$ original samples: $[\mathbf x^o, y^o]=\{(\mathbf{x}_i^o, y_i^o)\}_{i=1}^{{N/2}}$, where $y_i^o \in [1,C]$ and $N/2$ corresponding anonymized samples, and $[\mathbf x^a, y^a] =\{(\mathbf{x}_i^a, y_i^o + C) \}_{i=(N/2)+1}^{{N}}$.
Therefore, the number of speakers is $2C$ during the training of an OHNN-based anonymizer. 

We now explain the loss functions for learning the best values of $\Theta$ and $\Psi$ as defined in Eq.~(\ref{eqn:general_objective}). 
For the classification loss $\mathcal{L}_c$, we first consider the widely used AAM softmax loss \cite{deng2019arcface,xiang2019margin}:

\begin{equation}
        \mathcal{L}_{c} = \mathcal L_\text{AAM-softmax}  = -\frac{1}{N}\sum_{i=1}^{N}{{\rm log}\frac{e^{\left\|\mathbf w_{y_i}\right\| \cdot \left\|\mathbf x_i\right\|\cdot cos(\theta_{{y_i},i} +m_1 )}}{Z}}, 
            \label{eq:aam1}
    \end{equation}
where 

\begin{align*}
 Z=e^{||\mathbf w_{y_i}||\cdot||\mathbf{x}_i|| \cdot \cos(\theta_{{y_i},i} +m_1)} + \sum_{j=1, j \neq y_i}^{2C} e^{||\mathbf w_{j}||\cdot||\mathbf{x}_i||\cdot \cos(\theta_{{j},i})},
 \end{align*}
$\mathbf w_j$ is the $j$-th column of the weight in the fully-connected layer before the softmax layer, where $\mathbf w \in {\mathbb{R}^{d \times 2C}} $; and $\theta_{{y_i},i}$ is the angle between $\mathbf x_i$ and the target class's weight vector $\mathbf w_{y_i}$. 
After fixing the weight $||\mathbf w_{y_i}|| = 1 $ by $\ell_2$-normalization and rescaling $||\mathbf x_i||$ to $s$ to ensure that the gradient is not too small during training, we can write Eq.~(\ref{eq:aam1}) as
\begin{align}
   \mathcal{L}_{c} = \mathcal L_\text{AAM-softmax} = -\frac{1}{N}\sum_{i=1}^N\log \frac{e^{s({\cos(\theta_{{y_i}, i}+m_1)})}}{Z},
\label{eq:aam2}
\end{align}
where $ Z = {e^{s({\cos(\theta_{{y_i}, i}+m_1))}} + \sum_{j=1,j\neq y_i}^{2C} e^{s({\cos(\theta_{{j}, i})})}}$. Since the target label $y_i$ varies across the original and anonymized speakers, the classification loss $\mathcal{L}_\text{AAM-softmax}$ encourages the OHNN-based anonymizer to produce anonymized vectors that are varied for different speakers and distinct from original speaker vectors.

To further improve the discrepancy for original-anonymized (or anonymized-original) pair samples, we add an extra margin penalty $m_2$ into the AAM softmax loss.
The approach is called weighted additive angular margin (w-AAM) softmax. Let $i \in [1, N]$ be the index of the original (or anonymized) sample in a mini-batch based on this batch data construction method. The corresponding anonymized (or original) sample is indexed by $(i+N/2)\%N$, 
where $\%$ denotes the modulo operation. 
The proposed w-AAM-based loss function $\mathcal{L}_{\text{w-AAM}}$ is similar to $\mathcal{L}_{\text{w-AAM}}$ except that the factor $Z$ is defined as
\begin{equation}
\begin{aligned}
     Z = &e^{s({\cos(\theta_{{y_i}, i}+m_1))}} + {e^{s({\cos(\theta_{{y_{(i+N/2) \% N}}, i}-m_2))}}} \\ 
 &+
\sum_{j=1,j\neq i,j\neq (i+N/2) \% N }^{2C} e^{s({\cos(\theta_{{j}, i})})}.
\end{aligned}
\end{equation}

In our experiments, we set $m_1 = m_2 = 0.2$, $s=30$ and compared the performance with settings of $\mathcal L_c = \mathcal L_\text{AAM}$ and $\mathcal L_c  = \mathcal L_\text{w-AAM}$.

For the similarity metric $\mathcal{L}_s$, we choose the cosine similarity\footnote{\url{https://pytorch.org/docs/stable/generated/torch.nn.CosineEmbeddingLoss.html}} given by $\mathcal L_s (\mathbf{x}_i^o, \mathbf{x}_i^a) = \max(0,\cos( \mathbf{x}_i^o, \mathbf{x}_i^a) - m)$, we set the margin $m=0$. 
The cosine similarity is a reasonable choice because it is closer to what most ASV systems use for scoring the similarity between speaker vectors. As the anonymizers are trained to minimize the cosine similarity between original and anonymized speaker vectors, the anonymized speech is expected to be judged as a different speaker by the attacker ASV, hence protecting the speaker's identity. 

\section{Evaluation}
\label{sec:eval}
To evaluate the effectiveness of the SSL-based SAS using the proposed OHNN-based anonymizer under all the attack scenarios for English speaker anonymization, we followed the VPC evaluation plan \cite{tomashenko2020introducing,tomashenko2021voiceprivacy,tomashenko2022voiceprivacy} described in Section \ref{sec:related work}.
Then, we conducted anonymization experiments under a language-mismatched condition, using Mandarin data as the non-included language in the training database. The purpose of these experiments was to determine whether that the proposed OHNN-based anonymizer, which eliminates the need for an English speaker pool, can effectively reduce the language mismatch present in anonymized speaker representations. As a result, better speech content preservation is achieved for Mandarin speaker anonymization.

\begin{figure}
\centering
\subfloat[][{S-Select}]{\includegraphics[height=4cm]{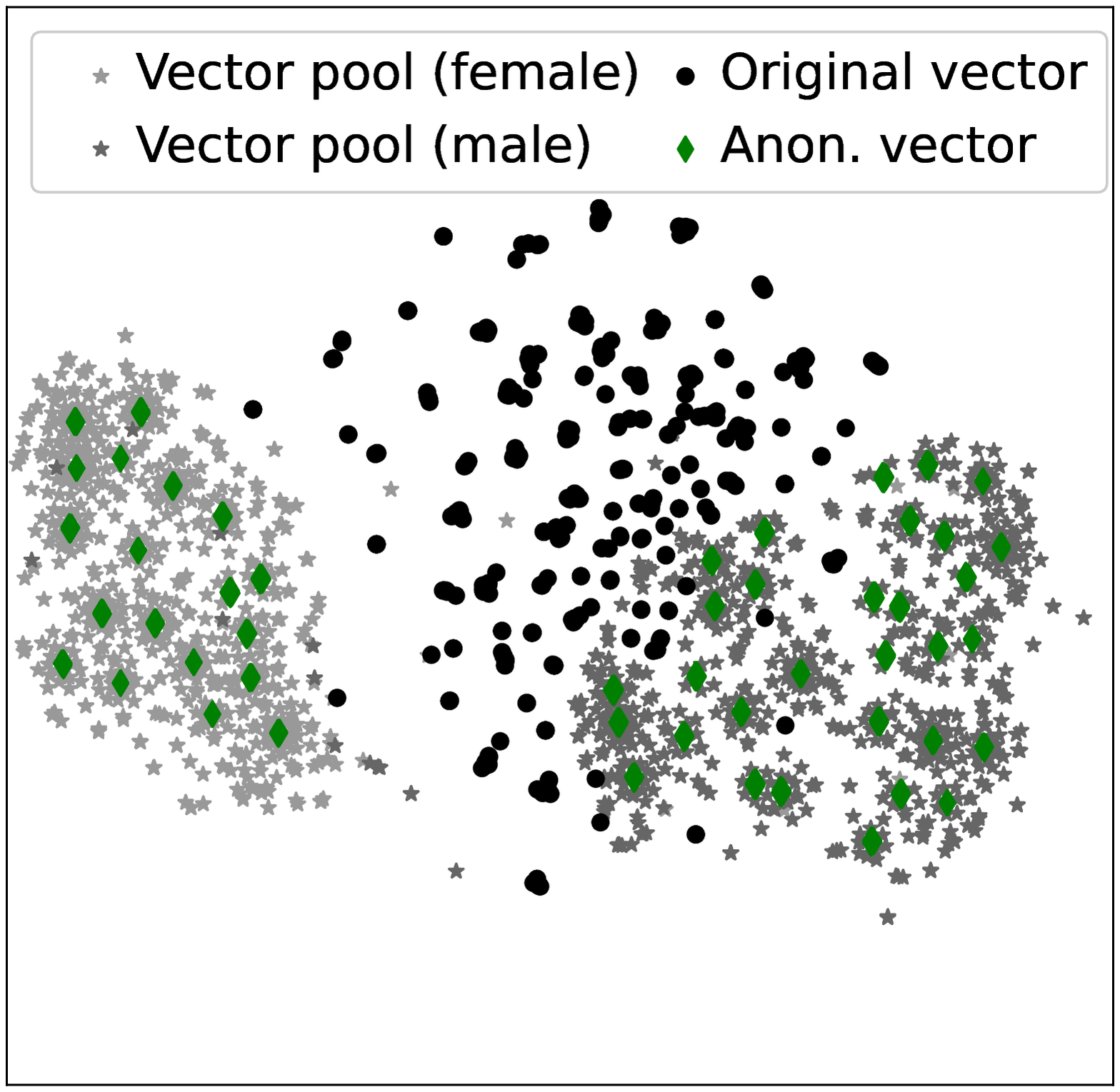}}
\hspace{0.5cm}
\subfloat[][S-LOH*]{\includegraphics[height=4cm]{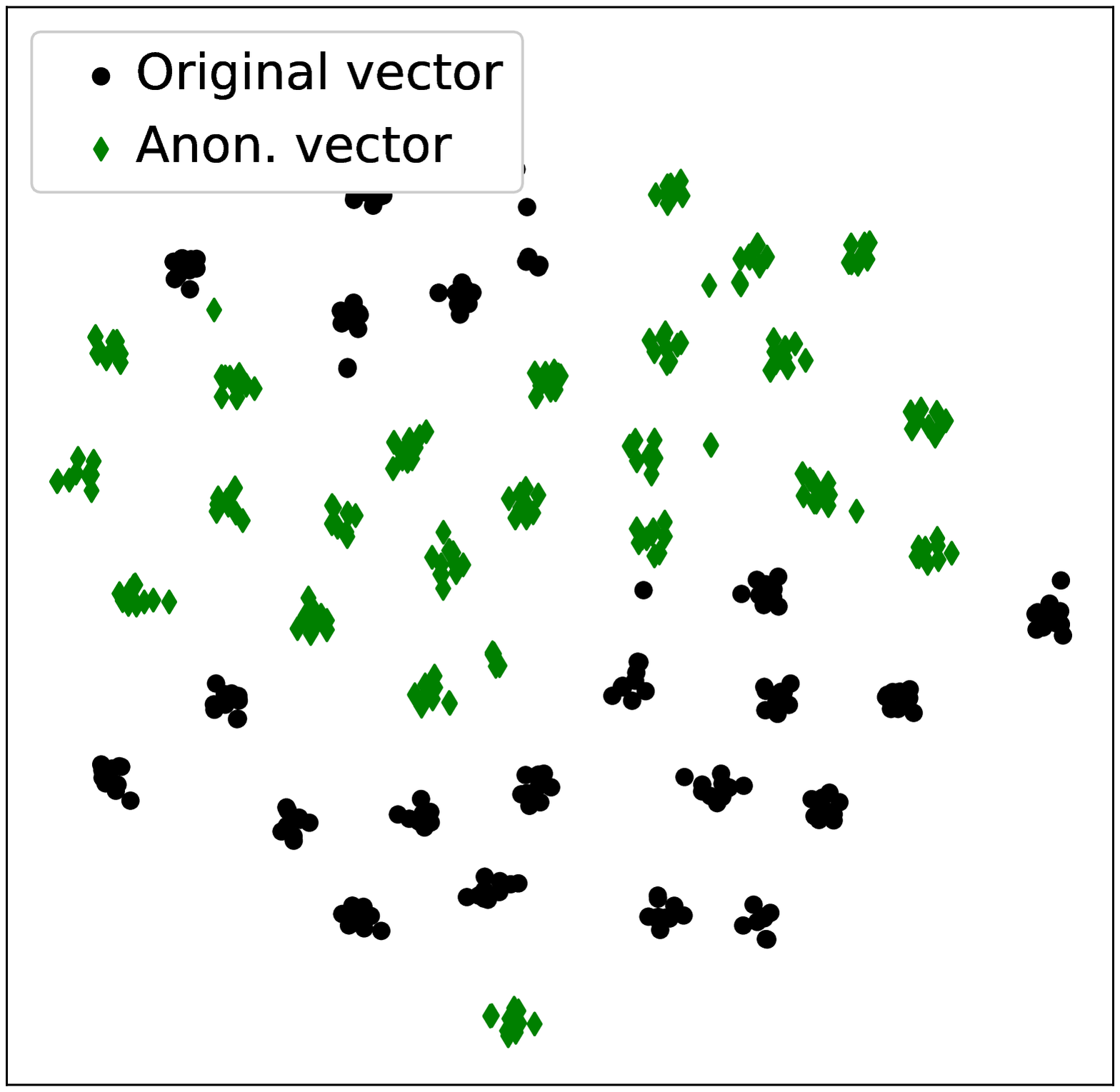}}\vspace{0.0cm}\\
\caption{Visualization of original and anonymized speaker vectors generated by the \textbf{S-Select} and \textbf{S-LOH*} anonymizers. }
\label{fig:compare_trans}
\end{figure}

\begin{figure*}[t]
\centering
\includegraphics[width=2\columnwidth]{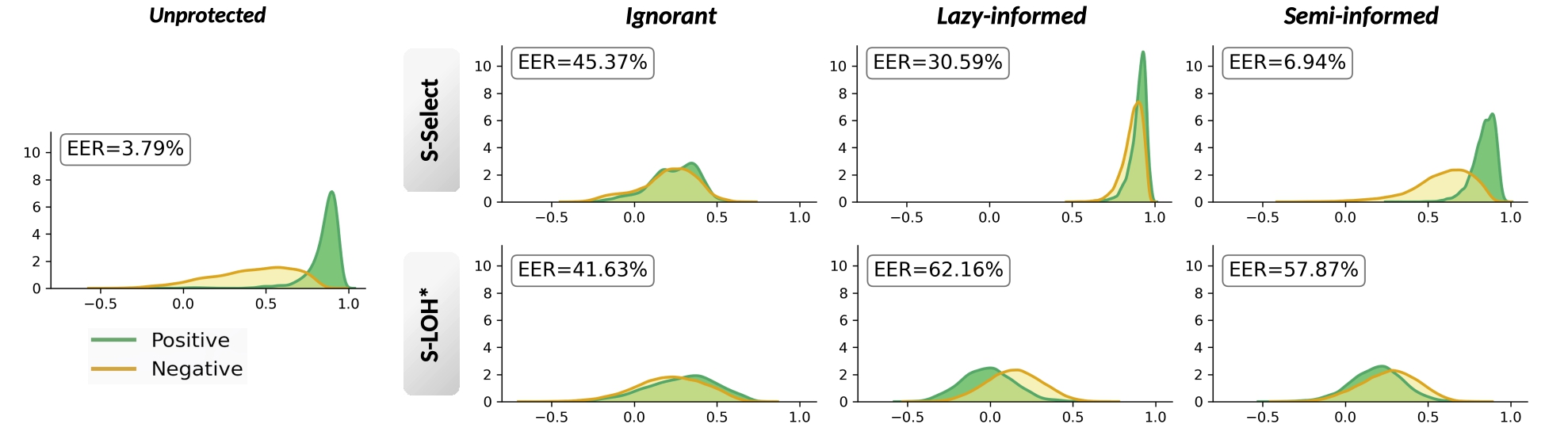}
\caption{Cosine similarities between pairs of the speaker vectors extracted from the generated speech of users and different attackers.
Positive: paired utterances from the same speaker. Negative: paired utterances from different speakers.}
\label{fig:cos_compare}
\end{figure*}

\subsection{Speaker Anonymization Dataset and Experimental Setup}
\subsubsection{Dataset}
The SSL-based SAS was built using the following VPC standard datasets \cite{tomashenko2020introducing}:
an ECAPA-TDNN speaker encoder trained on the \textit{VoxCeleb-2} \cite{chung2018voxceleb2};
a HuBERT-based soft content encoder finetuned from a pretrained HuBERT Base model\footnote{\url{https://github.com/pytorch/fairseq/tree/main/examples/hubert}} on \textit{LibriTTS-train-clean-100} \cite{zen2019libritts} to capture content representations; and a HiFi-GAN model trained on \textit{LibriTTS-train-clean-100} \cite{zen2019libritts}.

Unlike the selection-based anonymizer, which relies on an additional multi-speaker English dataset (\textit{LibriTTS-train-other-500}) containing data from 1,160 speakers as the external pool, the OHNN-based anonymizers reuse a multi-speaker multi-language dataset (\textit{VoxCeleb-2}), that is used to train the ECAPA-TDNN of the SSL-based SAS \cite{miao22_odyssey}.
This large-scale dataset contains over 1 million utterances by 5,994 speakers of 145 different nationalities.

English speaker anonymization was evaluated on the official VPC development and test sets \cite{tomashenko2020introducing,tomashenko2021voiceprivacy,tomashenko2022voiceprivacy}.
These two sets contain English utterances by several female and male speakers from the \textit{LibriSpeech} and \textit{VCTK} \cite{yamagishi2019cstr} corpora.
For the \textit{Ignorant} and \textit{Lazy-informed} conditions, we used the language-matched $ASV_\text{eval}$ system provided by the VPC  \cite{tomashenko2020introducing,tomashenko2021voiceprivacy,tomashenko2022voiceprivacy}.
It was trained on the original \textit{LibriSpeech-train-clean-360} English dataset.
For the \textit{Semi-informed} condition, we trained $ASV_\text{eval}^\text{anon}$ system in the same way as $ASV_\text{eval}$, but with anonymized speech data. Likewise, $ASR_\text{eval}$ and $ASR_\text{eval}^\text{anon}$ were trained with the same original and anonymized speech data, respectively.

The same anonymization systems used for English speakers were directly adopted for Mandarin speaker anonymization without training or fine-tuning on Mandarin data.
The evaluation for Mandarin was conducted on a test set sampled from a 20-hour, multi-speaker Mandarin corpus called \textit{AISHELL-3} \cite{shi21c_interspeech}. 
The test set contains 4,267 utterances by 44 speakers. We split the utterances into test trial (88 utterances) and enrollment (4,179 utterances) subsets, which were used to produce 10,120 enrollment-test pairs for ASV evaluation, including 2,200 same-speaker and 7,920 different-speaker pairs.
The ASV evaluation model {$ASV_\text{eval}^{\textrm{mand}}$}  under the \textit{Lazy-informed} condition was an ECAPA-TDNN trained on the Mandarin datasets  \textit{CN-Celeb-1 \& 2} \cite{li2022cn,fan2020cn}.
The ASV evaluation model under the \textit{Semi-informed} condition called $ASV_\text{eval}^{{\text{anon}}^\textrm{mand}}$ was fine-tuned from $ASV_\text{eval}^{\textrm{mand}}$ using anonymized utterances from 285 speakers in the interview, speech and live broadcasting genres of \textit{CN-Celeb-1 \& 2}.
The ASR evaluation model {$ASR_\text{eval}^{\textrm{mand}}$} 
was a publicly available ASR Transformer  \cite{speechbrain} trained on a 150-hour Mandarin ASR dataset, \textit{AISHELL-1} \cite{aishell_2017}. 

\begin{table}[t]
\centering
\caption{ Notations for the evaluated speaker anonymization methods.}
\label{tab:notations}
\resizebox{1\linewidth}{!}{
\begin{tabular}{clcccc}
\toprule
 & Notation & \makecell[c] {Content \\ encoder} &  \makecell[c]{Speaker  \\encoder}  & \makecell[c]{Syn.\\ model}   & \makecell[c] {Speaker \\ anon.} \\\midrule

\multirow{5}{*}{\rotatebox{90}{Disentangle}}&  {\textbf{B1.a} \cite{tomashenko2022voiceprivacy}}               & {F-TDNN}  & {TDNN}   &{SS-AM+NSF}  & {Select.}   \\  
  &  {\textbf{B1.b} \cite{tomashenko2022voiceprivacy}}               & {F-TDNN}  & {TDNN}   &{HiFi-GAN+NSF}  & {Select.}   \\ 

               &     {\textbf{S-Select} \cite{miao22_odyssey}}               & {SSL}  & {ECAPA}   & {HiFi-GAN}  & {Select.}   \\ 

                &       \textbf{S-ROH}              & {SSL}  & {ECAPA}   & {HiFi-GAN}  &   \makecell[c]{ROH}    \\     
                &       \textbf{S-LOH}               & {SSL}  & {ECAPA}   & {HiFi-GAN}  &  \makecell[c]{LOH}   \\             
\midrule
DSP & \textbf{B2}  \cite{tomashenko2022voiceprivacy}  &  \multicolumn{4}{c}{McAdams coefficients-based} \\
\bottomrule
\end{tabular}
}
\end{table}

\subsubsection{Experimental Setup}
Table \ref{tab:notations} lists notations for the different speaker anonymization approaches that we examined.
\textbf{B1.a}, \textbf{B1.b}, and \textbf{B2} are the baseline systems from VPC 2022 \cite{tomashenko2022voiceprivacy}.
\textbf{S-Select} denotes the SSL-based SAS using a selection-based anonymizer.
\textbf{S-ROH} denotes a system obtained by replacing the selection-based anonymizer of \textbf{S-Select} with a random OH (ROH) anonymizer and keeping other components unchanged.
Likewise, \textbf{S-LOH} indicates the use of a learnable OH (LOH) anonymizer.
Noted that, hereafter, \textbf{S-ROH*}  and \textbf{S-LOH*} refer to models trained with the w-AAM and cosine similarity losses.

For \textbf{S-Select}, the YAAPT algorithm \cite{kasi2002yet} is used to extract the F0.
The ECAPA-TDNN with 512 channels in the convolution frame layers \cite{desplanques2020ecapa} provides 192-dimensional speaker identity representations. 
The HuBERT-based soft content encoder \cite{van2021comparison} takes the CNN encoder and the first and sixth transformer layers of the pretrained HuBERT base model as a backbone. It downsamples a raw audio signal into a 768-dimensional continuous representation, which is then mapped to a 200-dimensional vector by one projection layer to predict discrete speech units.
These speech units are obtained by discretizing the intermediate 768-dimensional representations via \textit{k}-means clustering\footnote{\url{https://github.com/pytorch/fairseq/tree/main/examples/textless_nlp/gslm/speech2unit}}  \cite{polyak2021speech,lakhotia2021generative}.
The training procedures are detailed in~\cite{miao22_odyssey}.
For the selection-based anonymizer, attackers had different random seeds from users when randomly choosing 100 speaker vectors from the 200 farthest ones; thus, the attackers had different pseudo-speaker vectors.

The OHNN-based anonymizer accepts 192-dimensional speaker representations extracted from a pretrained ECAPA-TDNN, which was the same here as the ECAPA-TDNN of the SSL-based SAS. 
We followed the VPC evaluation plan, in which attackers in the \textit{Lazy-informed} and \textit{Semi-informed} scenarios have partial knowledge of the speaker anonymizer. They are assumed to know the training dataset, structure, loss functions, and other training parameters of the user's OHNN-based anonymizer, except for the training seed to initialize the training weights. 
Specifically, the training seeds were 50 and 1986 for users and attackers, respectively\footnote{The training seeds can be any values as long as they are different for users and attackers.}.
Using knowledge of the OHNN-based anonymizer, an attacker trains a new anonymizer to anonymize speech. 
All the OHNN-based anonymizers were trained with a cyclical learning rate \cite{smith2017cyclical}, which varied between 1e-8 and 1e-3, and the Adam optimizer \cite{kingma2014adam} by using the SpeechBrain \cite{speechbrain} toolkit based on PyTorch \cite{paszke2019pytorch}. The number of iterations of one cycle was set to 130k.
We fixed $d=192$, $L=12$ for both the ROH and LOH anonymizers,
but we use $q_l =192$ and $q_l=50$ for the ROH and LOH training, respectively.
The hyperparameter $\lambda$ in Eq.~(\ref{eqn:general_objective}) was set to 20 \footnote{Audio samples are available at \url{https://github.com/nii-yamagishilab/SSL-SAS}}.

\begin{figure}[t]
\centering
\includegraphics[width=0.9\columnwidth]{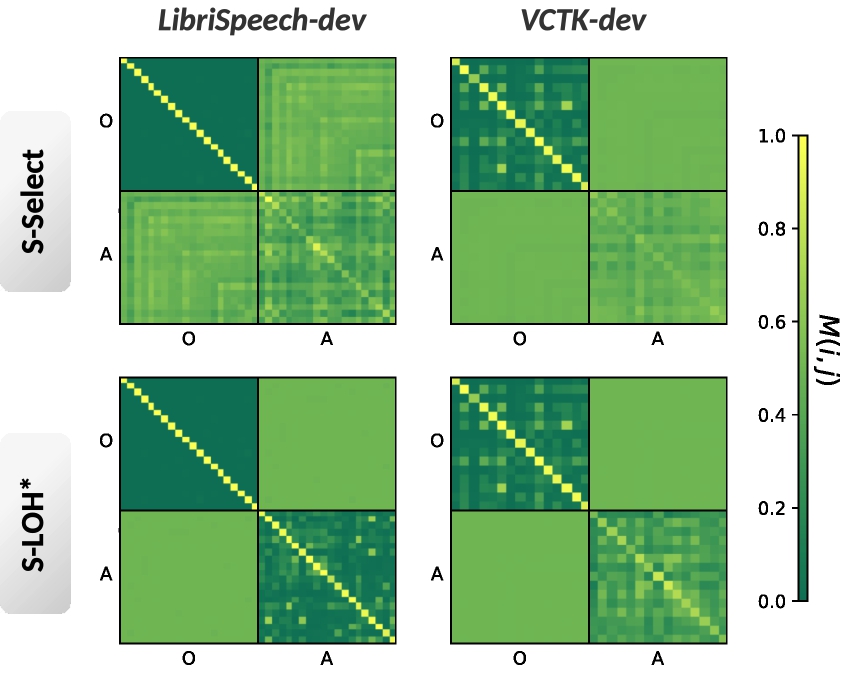}
\caption{Voice similarity matrices for \textbf{S-Select} and \textbf{S-LOH*} on the female speakers in the LibriSpeech-dev and VCTK-dev datasets. 
The global matrix $M$ for each system comprises three submatrices $M_\text{oo}$,  $M_\text{oa}$, and $M_\text{aa}$ defined in Section \ref{sec:evaluation} via
 {$M=\tiny\begin{pmatrix}
M_\text{oo} & M_\text{oa} \\
M_\text{oa} & M_\text{aa}
\end{pmatrix}$.} }
\label{fig:compare_vd}
\vspace{-2mm}
\end{figure}

\subsection{Speaker Anonymization Experiments in English}
For the English experiments, first, we explored the difference between selection- and OHNN-based anonymizers by comparing the performance of \textbf{S-Select} and \textbf{S-LOH*}.
Then, we investigated different configurations for the OHNN-based anonymizer, including the losses and whether to explicitly use speaker information to optimize the Householder transformation. 
Finally, we compared SSL-based speaker anonymization using an OHNN-based anonymizer with other approaches, including the disentanglement- and DSP-based approaches.

\begin{table*}[thbp]
\caption{
Average EER (\%), WER (\%), and $G_{\text{VD}} (dB)$ on the VPC English development(dev) and test sets. The speaker vectors were anonymized by an OHNN-based anonymizer with AAM+cos or w-AAM+cos. $\uparrow$ indicates better performance with higher values, while $\downarrow$ indicates better performance with lower values.}
\label{tab:loss}
\setlength{\tabcolsep}{7pt}
 \centering
 \footnotesize
\begin{tabular}{lcccccccccc}
\toprule

 & \multicolumn{2}{c}{\multirow{3}{*}{Original} }     & \multicolumn{8}{c}{OHNN-based anonymizer}   \\ 
&                  & & \multicolumn{4}{c}{\textbf{S-ROH}}    &  \multicolumn{4}{c}{\textbf{S-LOH}}    \\
\cmidrule(lr){4-7} \cmidrule(lr){8-11} 
  &                  &   & \multicolumn{2}{c}{AAM+cos}  & \multicolumn{2}{c}{w-AAM+cos}    & \multicolumn{2}{c}{AAM+cos} & \multicolumn{2}{c}{w-AAM+cos}  \\
\cmidrule(lr){2-3} \cmidrule(lr){4-5}  \cmidrule(lr){6-7} \cmidrule(lr){8-9} \cmidrule(lr){10-11} 
    & \multicolumn{1}{c}{dev}  & test     & \multicolumn{1}{c}{dev}    & test    & \multicolumn{1}{c}{dev}    & test    & \multicolumn{1}{c}{dev}      & test     & \multicolumn{1}{c}{dev}     & test    \\ 
\textit{Ignorant} by  $ASV_\text{eval}$  $\uparrow$    & \multicolumn{1}{c}{3.54} & 3.79   &  \multicolumn{1}{c}{43.28} & 45.09 & \multicolumn{1}{c}{47.60} & 49.83 & \multicolumn{1}{c}{45.19}   & 42.17  & \multicolumn{1}{c}{45.94}  & 41.63 \\ 
\textit{Lazy-informed} {$ASV_\text{eval}$ }  $\uparrow$  & \multicolumn{1}{c}{{3.54}} &  {3.79}  & \multicolumn{1}{c}{40.20} & 47.37  & \multicolumn{1}{c}{41.69} & 45.16 & \multicolumn{1}{c}{47.49}   & 49.62  & \multicolumn{1}{c}{59.31}  & 62.16  \\ 
\textit{Semi-informed}  {$ASV_\text{eval}^{\text{anon}}$ }  $\uparrow$ & \multicolumn{1}{c}{-} & -   & \multicolumn{1}{c}{7.75}  & 42.41 & \multicolumn{1}{c}{41.42} & 41.88 & \multicolumn{1}{c}{41.79}   & 40.66   & \multicolumn{1}{c}{60.12}  & 57.87 \\
\midrule
WER  by  {$ASR_\text{eval}$ } $\downarrow$  & \multicolumn{1}{c}{7.30} & 8.48  &  \multicolumn{1}{c}{9.31} & 10.20  & \multicolumn{1}{c}{9.32} & 10.24 & \multicolumn{1}{c}{9.31}   & 10.22  & \multicolumn{1}{c}{9.28}  & 10.20  \\ 
WER   by  {$ASR_\text{eval}^{\text{anon}}$}  $\downarrow$  & \multicolumn{1}{c}{-} & -  &  \multicolumn{1}{c}{7.52} & 7.72 & \multicolumn{1}{c}{7.67} & 7.84 & \multicolumn{1}{c}{7.47}   &  7.99 & \multicolumn{1}{c}{7.52}  & 7.94 \\ 
\midrule
$G_{\text{VD}}$ $\uparrow$ & \multicolumn{1}{c}{0} & 0  &  \multicolumn{1}{c}{-1.86} & -1.59 & \multicolumn{1}{c}{-1.92} & -1.64 & \multicolumn{1}{c}{-3.89}   & -3.55  & \multicolumn{1}{c}{-2.52}  & -2.25 \\ 
\bottomrule
\end{tabular}
\end{table*}

\subsubsection{Comparison of Selection- and OHNN-Based Anonymizers}
\label{sec:compare_anonymizer}
In the first experiments, we visualized the original and anonymized speech generated by \textbf{S-Select} and \textbf{S-LOH*} in terms of speaker embeddings, the cosine similarity of the speech pairs, and voice distinctiveness.

\noindent
\textbf{Original and anonymized speaker embeddings}:
To show the difference between the \textbf{S-Select} and \textbf{S-LOH*} anonymizers, we first applied t-distributed stochastic neighbor embedding (t-SNE)  \cite{van2008visualizing} to visualize the original and anonymized embeddings. The results are shown in Fig. \ref{fig:compare_trans}. The speaker embeddings were extracted from 50 speakers in the \textit{VoxCeleb-2} training set, which are shown in different colors, and 10 utterances were randomly selected from each speaker. Clearly, the anonymized speaker vectors generated by \textbf{S-Select} were heavily dependent on the distribution of an external pool, whereas \textbf{S-LOH*} generated distinctive anonymized speaker vectors that followed the distribution of the original speaker vector space. 

\noindent
\textbf{Cosine similarity distribution on speech pairs}:
Fig. \ref{fig:cos_compare} plots the cosine similarities between pairs of speaker vectors extracted from generated speech for all the test sets of \textit{LibriSpeech} and \textit{VCTK} on speech pairs provided by \cite{tomashenko2021voiceprivacy}. 
Depending on the attack condition, the speech can be original or anonymized generated by \textbf{S-Select} or \textbf{S-LOH*}.
For the \textit{Unprotected} condition, shown on the left side of Fig. \ref{fig:cos_compare}, the positive cosine similarity distributions (green) are close to 1, 
and the negative distributions (yellow) are close to 0, which indicates that the speaker vectors of the original speech were highly discriminative.
To protect speaker privacy, an ideal SAS should push the positive score distributions toward the negative ones regardless of the attacker type.

On the right side of Fig. \ref{fig:cos_compare}, the top part shows the score distributions for three attacker conditions with \textbf{S-Select}. There are much bigger overlaps of the positive and negative distributions for the \textit{Ignorant} condition than for the \textit{Unprotected} condition, which means that \textbf{S-Select} achieved reasonable speaker privacy performance under the \textit{Ignorant} condition.
Unfortunately, the overlaps are smaller for the \textit{Lazy-informed} and \textit{Semi-informed} conditions. 
This reveals the reason for the significant speaker privacy leakage under more powerful attack conditions.
Moreover, most of the cosine similarity scores are very close to 1, which may pose a risk of reducing the diversity of the anonymized speakers. 

The bottom right of Fig. \ref{fig:cos_compare} shows the score distributions for three attacker conditions with \textbf{S-LOH*}. The overlaps of the positive and negative distributions are well magnified under all the attack scenarios.
This verifies the effectiveness of our OHNN-based anonymizer in ensuring that the attackers cannot gain significant speaker privacy information from users.
Furthermore, most of the cosine similarity scores are far from 1, indicating the diversity of the anonymized speakers.

\begin{table*}[t]
\caption{
 Average EER (\%), WER (\%), and  $G_{\text{VD}}(dB)$ on the VPC English dev and test sets when processed by various speaker anonymization systems.}
\label{tab:baseline-compare-selection-based}
\setlength{\tabcolsep}{6pt}
  \centering
  \footnotesize
  \begin{tabular}{lcccccccccccccc}  
\toprule
&  \multicolumn{2}{c}{} &    \multicolumn{2}{c}{DSP} &  \multicolumn{6}{c}{\makecell[c]{ Selection-based anonymizer}}     &     \multicolumn{4}{c}{\makecell[c]{OHNN-based anonymizer}}    \\  \cmidrule(lr){6-11}  \cmidrule(lr){12-15} 
  & \multicolumn{2}{c}{Original}   & \multicolumn{2}{c}{\textbf{B2} \cite{tomashenko2022voiceprivacy}} & \multicolumn{2}{c}{ \textbf{B1.a} \cite{tomashenko2022voiceprivacy}}    & \multicolumn{2}{c}{\textbf{B1.b} \cite{tomashenko2022voiceprivacy}}      & \multicolumn{2}{c}{\textbf{S-Select} \cite{miao22_odyssey}} & \multicolumn{2}{c}{\textbf{S-ROH*} } & \multicolumn{2}{c}{\textbf{S-LOH*}  }  
\\  \cmidrule(lr){2-3} \cmidrule(lr){4-5}  \cmidrule(lr){6-7} \cmidrule(lr){8-9}  \cmidrule(lr){10-11}   \cmidrule(lr){12-13}  \cmidrule(lr){14-15} 

    & \multicolumn{1}{c}{dev}  & test     & \multicolumn{1}{c}{dev}  & test   & \multicolumn{1}{c}{dev}    & test    & \multicolumn{1}{c}{dev}    & test  & \multicolumn{1}{c}{dev}    & test        & \multicolumn{1}{c}{dev}    & test  & \multicolumn{1}{c}{dev}    & test   \\ 

\textit{Ignorant} by  {$ASV_\text{eval}$ $\uparrow$ }    & \multicolumn{1}{c}{3.54} & 3.79 & \multicolumn{1}{c}{37.01} & 38.29  & \multicolumn{1}{c}{53.14} & 50.29   & \multicolumn{1}{c}{53.91} & 52.14  &\multicolumn{1}{c}{48.23} & 45.37 & \multicolumn{1}{c}{47.60} & 49.83 & \multicolumn{1}{c}{45.94} & 41.63 \\ 
\textit{Lazy-informed}  {$ASV_\text{eval}$ $\uparrow$ }  & \multicolumn{1}{c}{3.54} & 3.79  & \multicolumn{1}{c}{43.80} & 45.04  & \multicolumn{1}{c}{32.12} & 32.82 & \multicolumn{1}{c}{27.39} & 27.51  & \multicolumn{1}{c}{29.29} & 30.59 & \multicolumn{1}{c}{41.69} & 45.16 & \multicolumn{1}{c}{59.31} & 62.16\\ 
\textit{Semi-informed}  {$ASV_\text{eval}^{\text{anon}}$ $\uparrow$ } & \multicolumn{1}{c}{-} & -   &\multicolumn{1}{c}{6.53} &  7.77  & \multicolumn{1}{c}{11.74} & 11.81 & \multicolumn{1}{c}{9.93} & 9.18 &  \multicolumn{1}{c}{7.75} & 6.94 & \multicolumn{1}{c}{41.42} & 41.86 & \multicolumn{1}{c}{60.12} & 57.87  \\
\midrule

WER  by  {$ASR_\text{eval}$ } $\downarrow$  & \multicolumn{1}{c}{7.30} & 8.48   & \multicolumn{1}{c}{17.15} & 18.52  & \multicolumn{1}{c}{10.88} & 10.98  & \multicolumn{1}{c}{10.69} & 10.84  & \multicolumn{1}{c}{8.73} & 9.77  & \multicolumn{1}{c}{9.32} & 10.24 & \multicolumn{1}{c}{9.28} & 10.20 \\ 
WER   by  {$ASR_\text{eval}^{\text{anon}}$} $\downarrow$  & \multicolumn{1}{c}{-} & -  & \multicolumn{1}{c}{8.04 } & 9.03    & \multicolumn{1}{c}{7.94} & 8.29 & \multicolumn{1}{c}{7.59} & 7.56 & \multicolumn{1}{c}{7.74} & 8.44 & \multicolumn{1}{c}{7.67} & 7.84 & \multicolumn{1}{c}{7.52} & 7.94 \\
\midrule
$G_{\text{VD}}$ $\uparrow$ & 0 & 0 & \multicolumn{1}{c}{-1.72 } & -1.63 & \multicolumn{1}{c}{-9.17} & -10.15 & \multicolumn{1}{c}{-6.44} & -6.44     & \multicolumn{1}{c}{-7.90} & -7.99 & \multicolumn{1}{c}{-1.92} & -1.64 & \multicolumn{1}{c}{-2.52} & -2.25 \\
\bottomrule 
\\[-1.8ex]
\multicolumn{6}{c}{}{* indicates that w-AAM+cos was used for training \textbf{S-ROH} and \textbf{S-LOH}.} 

\end{tabular}
\end{table*}

\begin{figure*}
\centering
\vspace{-0.2cm}
\subfloat[][B2]{\includegraphics[width=0.25\linewidth]{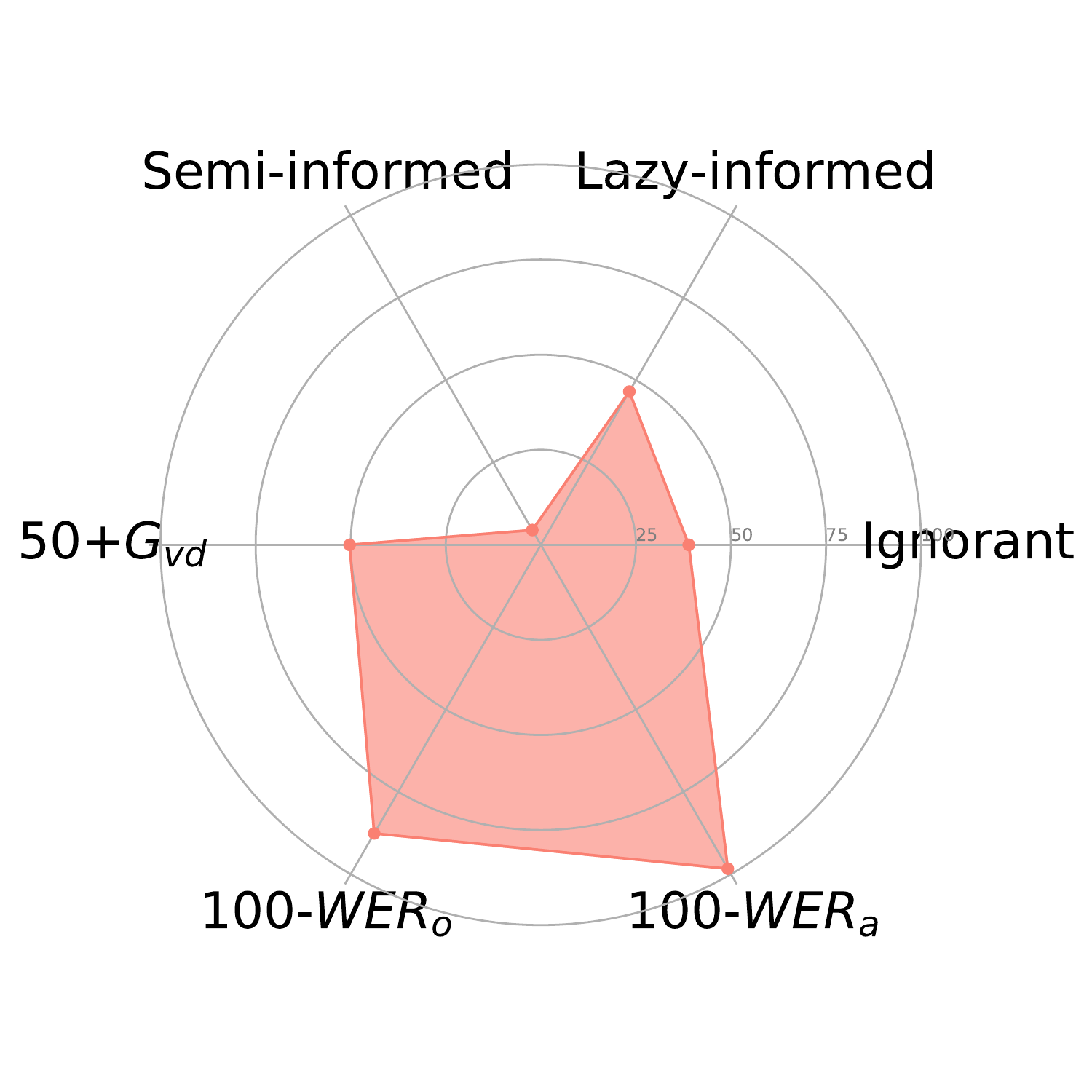}}
\subfloat[][B1.a]{\includegraphics[width=0.25\linewidth]{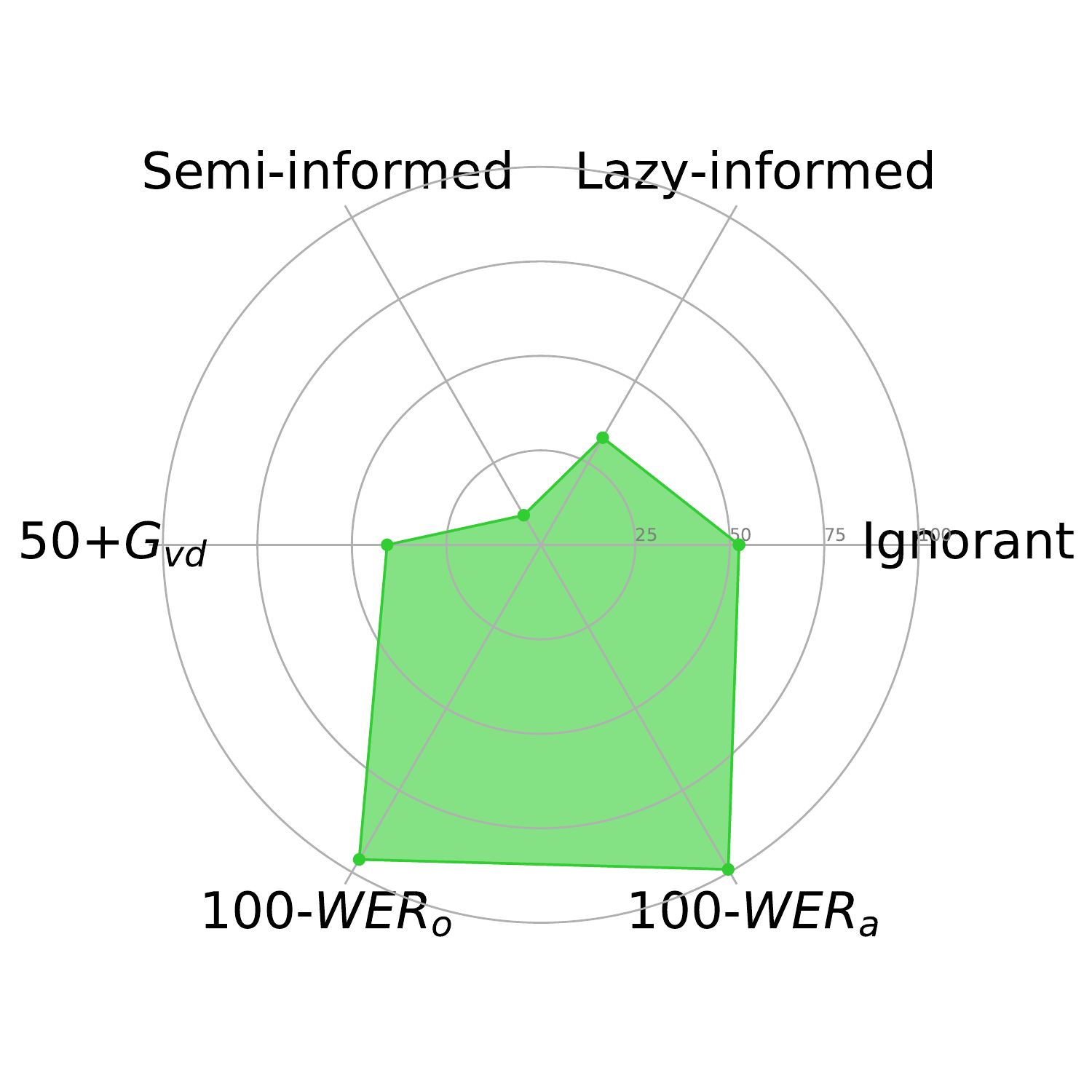}}
\subfloat[][B1.b]{\includegraphics[width=0.25\linewidth]{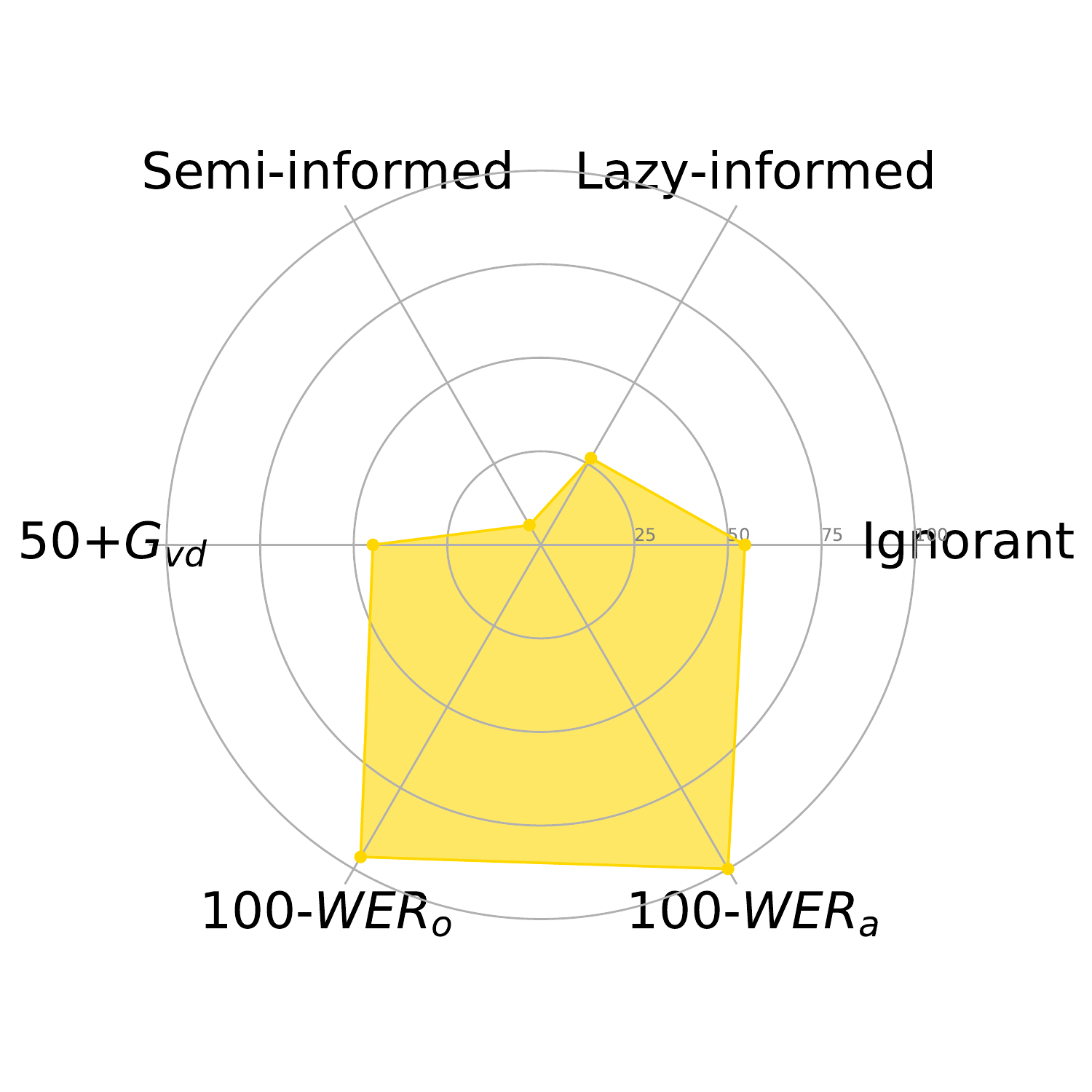}}
\vspace{-0.1cm}\\
\subfloat[][S-Select]{\includegraphics[width=0.25\linewidth]{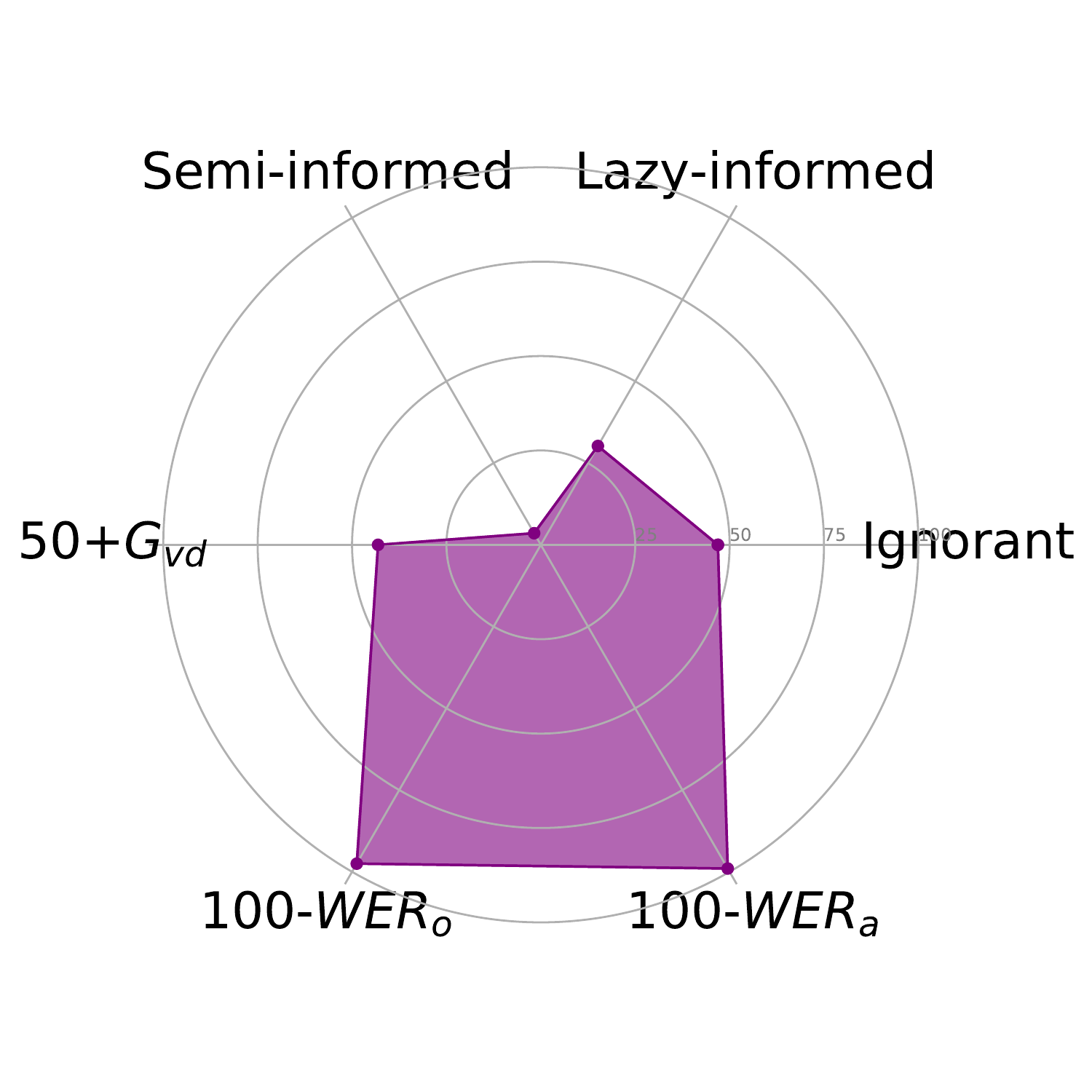}}
\subfloat[][S-ROH*]{\includegraphics[width=0.25\linewidth]{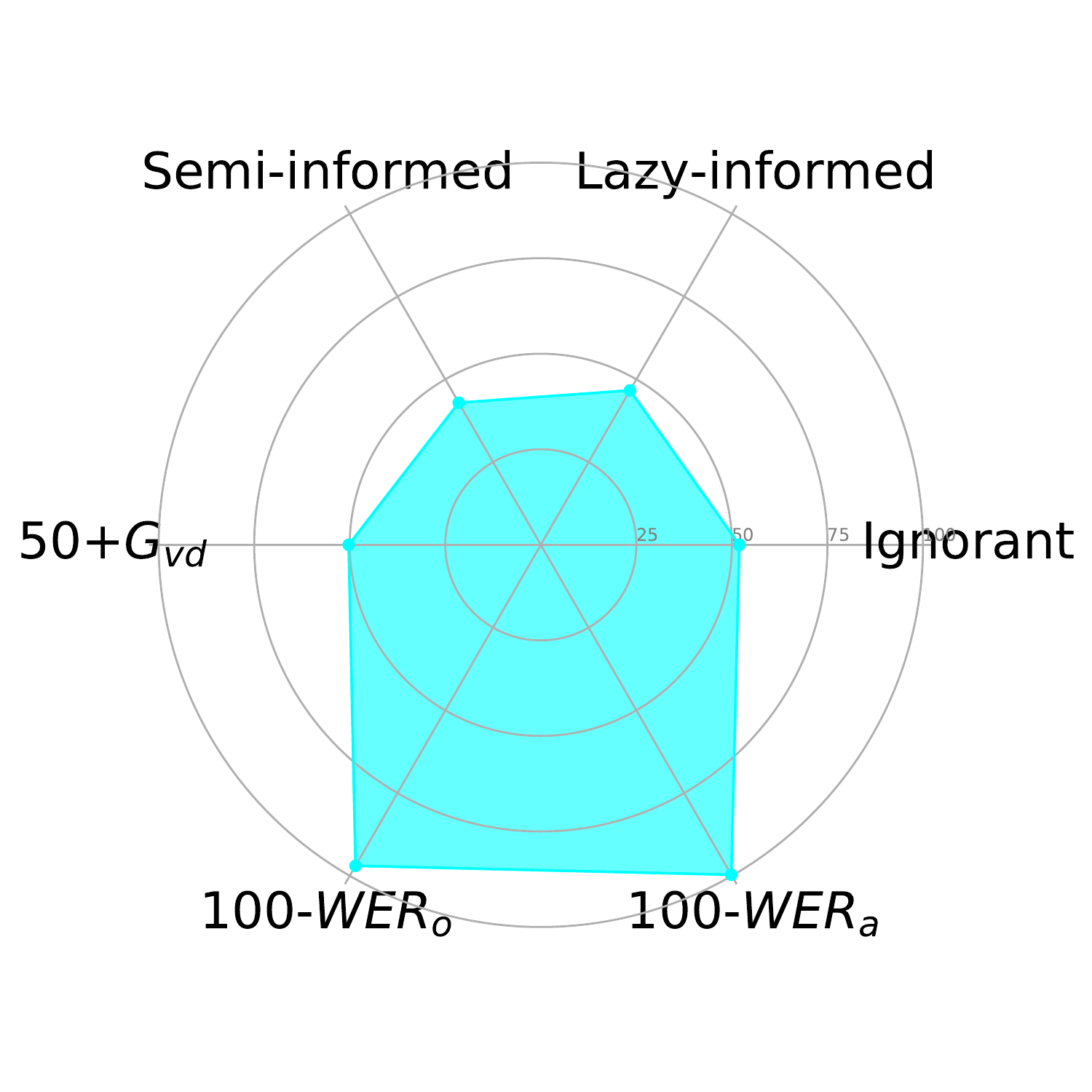}}
\subfloat[][S-LOH*]{\includegraphics[width=0.25\linewidth]{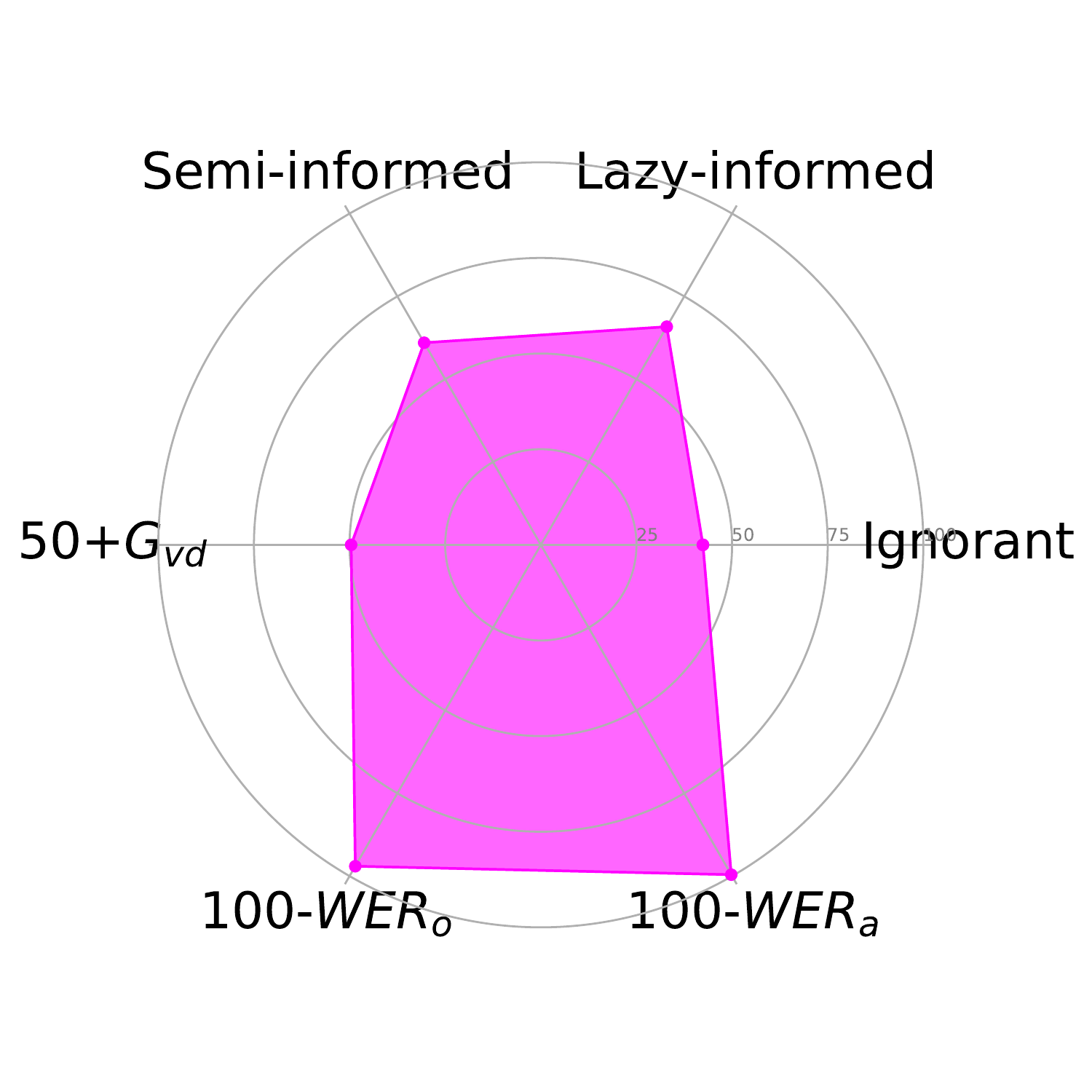}}

    \caption{ Radar charts for each system on English speaker anonymization. All values are rescaled to [0,100].}
    \label{fig:radar}
\end{figure*}
\noindent
\textbf{Comparison of gain of voice distinctiveness ($G_{\text{VD}}$)}:
Fig. \ref{fig:compare_vd} shows voice similarity matrices obtained for \textbf{S-Select} and \textbf{S-LOH*}.
The upper-left submatrix of each matrix $M$ is $M_\text{oo}$, and the distinct diagonal reflects the high voice distinctiveness within the original speech.
The upper-right (or lower-left) submatrix $M_\text{oa}$ reflects the voice similarity between the original and the anonymized speech, such that the diagonal disappears when they differ.
The lower-right submatrix $M_\text{aa}$ reflects the voice similarity within the anonymized speech, where a dominant diagonal appears if the anonymized speakers remain distinguishable \cite{noe2020speech}. 
There is a very weak dominant diagonal in $M_\text{aa}$ for \textbf{S-Select}, indicating that voice distinctiveness was lost among the anonymized speakers.
In contrast, the matrices for \textbf{S-LOH*} exhibit distinct diagonals in $M_\text{aa}$, indicating that voice distinctiveness was preserved after anonymization. 

In general, the \textbf{S-LOH*} anonymizer met the three constraints described in Section \ref{seq:problem_def}: good privacy protection, voice distinctiveness, and naturalness of the speaker vector space from the above analysis and visualization.

\subsubsection{Effects of Various Components for Proposed OHNN-Based Anonymizer}
The proposed OHNN-based anonymizer has two novel components: the loss functions and the Householder transformations.
Table \ref{tab:loss} summarizes the average EERs and WERs\footnote{The EER weights and detailed results for each subset are given in Appendix \ref{sec:app}. Due to limited space, other results are moved to the appendix of the paper on Arxiv.} under all attack scenarios using two OHNN-based anonymizers with different losses. In the table, $\uparrow$ indicates a better performance with higher values, while $\downarrow$ indicates a better performance with lower values.

\noindent
\textbf{Effect of the different losses}:
For the proposed OHNN-based anonymizer, w-AAM+cos performed better than AAM+cos in terms of the EER under most attacker conditions. This was because the introduced margin of w-AAM expands the inter-class variance of original-anonymized pairs, thus increasing the dissimilarity.

\noindent
\textbf{Effect of different Householder transformations}:
Clearly, the LOH anonymizers generally achieved better EERs than the ROH did. This result supports the view that, instead of using a global transformation for ROH, the LOH is more flexible because it learns from the speaker embeddings and thus brings more discriminative information.

For the WERs, those computed by ${ASR}_\text{eval}^\text{anon}$ were consistently lower than those of ${ASR}_\text{eval}$ for all systems.
This implies that such utility degradation due to OHNN-based anonymizers can easily be offset by training ASR evaluation models on similar anonymized data. 
Meanwhile, all the OHNN-based anonymizers achieved similar WERs with ${ASR}_\text{eval}$ or $ASR_\text{eval}^\text{anon}$, which confirms that the orthogonality of ROH and LOH did not change the distributions of the original and anonymized speaker vectors.

\begin{table*}[htb]
\centering
\caption{EER (\%) and CER (\%) on Mandarin data with {$ASV_\text{eval}^{\textrm{mand}}$},  $ASV_\text{eval}^{{\text{anon}}^\textrm{mand}}$, and {$ASR_\text{eval}^{\textrm{mand}}$}.
A higher EER indicates better privacy, while a lower CER indicates better intelligibility.}  
 \label{tab:mand_result}
\begin{tabular}{lcccccccc}
\toprule
\footnotesize
 & \multirow{3}{*}{Original }   &  DSP  & \multicolumn{2}{c}{ Selection-based anonymizer}    & \multicolumn{4}{c}{OHNN-based anonymizer}   \\ \cmidrule(lr){4-5}  \cmidrule(lr){6-7} \cmidrule(lr){8-9} 
&      &   \multirow{2}{*}{\textbf{B2} } &   \multirow{2}{*}{ \textbf{{B1.a}} }&  \multirow{2}{*}{\textbf{S-select}} &  \multicolumn{2}{c}{\textbf{S-ROH}}    & \multicolumn{2}{c}{\textbf{S-LOH} }\\  \cmidrule(lr){6-7} \cmidrule(lr){8-9} 
    &    &   &   & & AAM+cos & w-AAM+cos & AAM+cos & w-AAM+cos  \\ \midrule           
\textit{Ignorant} by {$ASV_\text{eval}^{\textrm{mand}}$} $\uparrow$  & 2.04 & 35.50 & 44.54& 37.90 & 32.09 &34.82 & 33.28 & 33.27 \\ 
\textit{Lazy-informed} by {$ASV_\text{eval}^{\textrm{mand}}$} $\uparrow$  & 2.04 & 36.31& 41.54 & 22.58 &34.04 & 34.54 & 39.54 & 47.49   \\
\textit{Semi-informed} by $ASV_\text{eval}^{{\text{anon}}^\textrm{mand}}$ $\uparrow$  & - &  42.73 & 42.44 & 19.82 &42.81 & 41.72 &  40.72 & 48.99   \\

\midrule
CER by {$ASR_\text{eval}^{\textrm{mand}}$} $\downarrow$  & 10.36 & 61.90& 68.67& 18.92 & 17.15 & 17.28 & 17.20 & 17.90  \\
\bottomrule
\end{tabular}
\end{table*}

\subsubsection{Comparison of Various SASs Using Different Anonymizers}

\noindent
\textbf {Primary privacy and utility evaluation}:
Table \ref{tab:baseline-compare-selection-based} lists the average EER and WER results for various SASs under all scenarios. 
To anonymize the speaker representations, \textbf{B2} randomly alters the formant position, \textbf{B1.a}, \textbf{B1.b}, and \textbf{S-Select} used the selection-based anonymizer,  while \textbf{S-ROH*} and \textbf{S-LOH*} used the OHNN-based anonymizer.

First, we examine the results with the selection-based anonymizer.
Using the selection-based anonymizer, the EERs of \textbf{S-Select},  \textbf{B1.a} and \textbf{B1.b} decreased by around 30\% under the \textit{Lazy-informed} condition and 7\%-9\% under the \textit{Semi-informed} condition, indicating severe speaker privacy leakage.

Next, we examine the results with the proposed OHNN-based anonymizer integrated into different configurations.
First, \textbf{S-ROH*} and \textbf{S-LOH*} could protect speaker information almost as well as the VPC baselines (\textbf{B1.a and \textbf{B1.b}}) could when facing the \textit{Ignorant} attacker. Moreover, for the \textit{Lazy-informed} and \textit{Semi-informed} attackers, it comfortably outperformed all the baseline systems, achieving over 40\% EER.
Second, among all the methods, \textbf{S-ROH*} and \textbf{S-LOH*} preserved speech content the best with {$ASR_\text{eval}^\text{anon}$}, achieving even lower WERs than for original speech on average.

Another interesting observation is that, while \textbf{B2}, \textbf{B1.a}, \textbf{B1.b}, and \textbf{S-select} are effective for protecting user privacy under the \textit{Ignorant} condition, the utility performance in terms of WER and $\text G_\text{VD}$ is worse than that of the OHNN-based anonymizers. This suggests that the baseline methods sacrifice utility to achieve a high privacy protection performance. Our proposed methods achieve a good balance between improving both privacy and utility metrics under various attack scenarios.

\noindent
\textbf {Secondary utility evaluation}:
The bottom of Table \ref{tab:baseline-compare-selection-based} lists the results for the average gain of voice distinctiveness, $\text G_\text{VD}$.
They indicate that our proposed \textbf{S-ROH*}  and \textbf{S-LOH*} achieved much better preservation of voice distinctiveness than the SASs using the selection-based anonymizer. 
The $\text G_\text{VD}$ results of the \textbf{S-select} and OHNN-based anonymizers again confirm the findings described in Section \ref{sec:compare_anonymizer}.

\begin{figure}[t]
\centering
\vspace{-5mm}
\includegraphics[width=0.9\columnwidth]{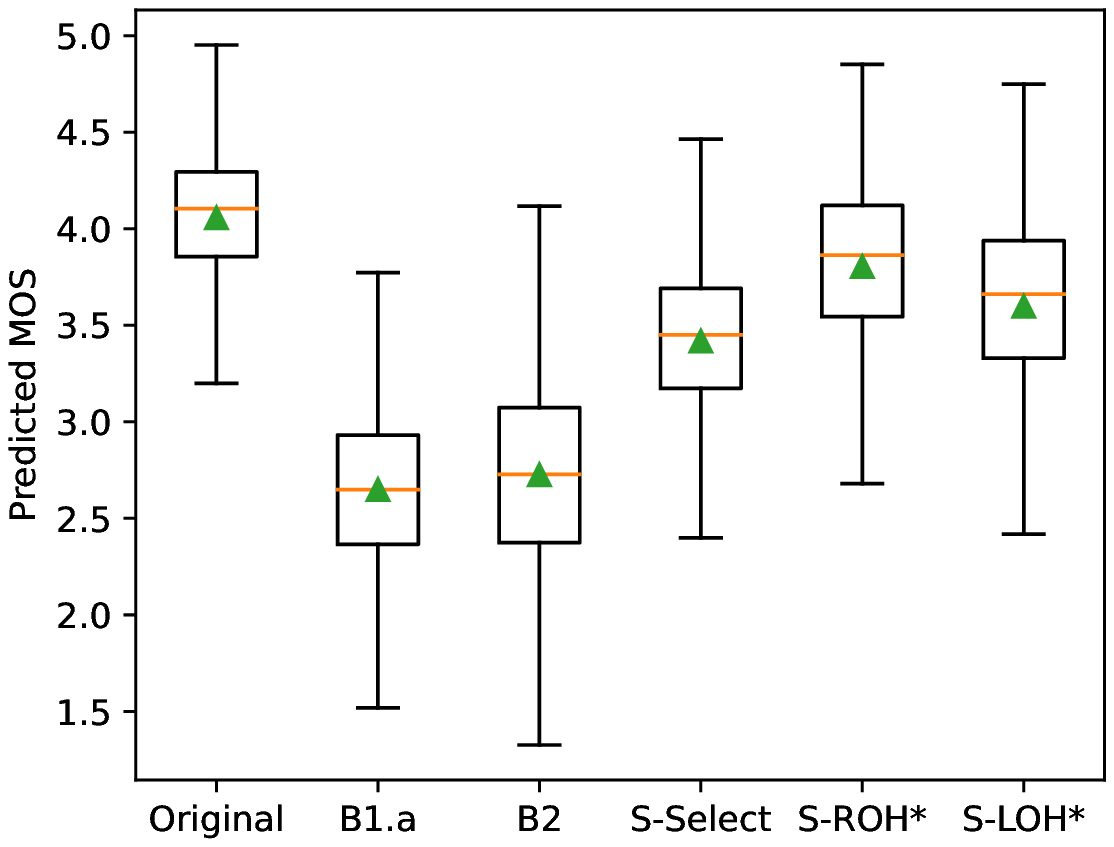}
\caption{Box plots on predicted naturalness scores of anonymized speech from experimental systems. Triangle symbols and the bar marks in the boxes represent mean and median scores, respectively.}
\label{fig:mos}
\vspace{-2mm}
\end{figure}

\noindent
\textbf {MOS prediction}: To further analyze the effectiveness of our proposed models,  we utilize a recently proposed mean opinion score (MOS) prediction network \cite{cooper2021generalization} to estimate the perceived naturalness as another utility metric. Box plots of the predicted MOS scores are shown in Fig.~\ref{fig:mos}. The results demonstrate that \textbf{S-Select} has a higher naturalness than \textbf{B2} and \textbf{B1.a}. After replacing the selection-based anonymizer with the OHNN-based anonymizers \textbf{S-ROH*} and \textbf{S-LOH*}, we see a further improvement in naturalness. 

Note that we used predicted MOS rather than human perception-based MOS obtained through listening tests in light of time and cost limits. The predicted MOS is reasonably well-aligned with human perception \cite{cooper2021generalization}. In Fig.~\ref{fig:mos}, we can see that the ranking of the predicted MOS of the original, \textbf{B1.a} , and \textbf{B2} are consistent with those from the listening test done by the VPC \cite{tomashenko2021voiceprivacy}.

\noindent
\textbf {Overall performance}:

As there are multiple metrics for evaluating the model performance, we summarize the results using a radar chart for each system in Fig.~\ref{fig:radar}. Each radar chart covers the EER values under the \textit{Ignorant}, \textit{Lazy-informed}, and \textit{Semi-informed} conditions, $WER_o$ by $ASR_\text{eval}$, $WER_a$ by $ASR_\text{eval}^{\text{anon}}$, and $G_{\text{VD}}$. Note that the chart shows $100-WER$, so the higher the better. Accordingly, a larger shaded area in the radar plot indicates a better overall performance. It is evident that the proposed \textbf{S-ROH*} and \textbf{S-LOH*} achieve larger shaded areas than the other systems, which performed particularly worse under the challenging semi-informed condition.

\subsection{Speaker Anonymization Experiments in Mandarin}

Table \ref{tab:mand_result} lists the EERs and CERs for the Mandarin test dataset. 
The first observation is that baselines \textbf{B1.a} and \textbf{B2} obtained EERs higher than 30\% under the three conditions, but the CERs were higher than 60\%. These results indicate that both systems achieved a high level of speaker identity protection by heavily distorting the speech contents. In particular, the results of \textbf{B1.a} suggest that it was inappropriate to use the ASR AM trained on the English data to extract speech content from the Mandarin data. 
The second observation is that the trends for the \textbf{S-Select} and OHNN-based anonymizers with different losses were remarkably similar to those observed on the English test sets.

The proposed OHNN-based anonymizers obtained ASV EERs higher than 30\% under all evaluation conditions, and the CERs were lower than those of other systems. Compared to the baselines, the proposed systems adequately protected the speaker information without heavily sacrificing the speech contents. Compared to the selection-based system, the proposed system[s?] achieved a lower CER while obtaining much higher ASV EERs, particularly in the most challenging \textit{Lazy-informed} and \textit{Semi-informed} scenarios. In particular, the CER on the anonymized speech decreased to less than 18\% with the OHNN-based anonymizers, suggesting improved utility.
One possible reason for the decreased CER when using OHNN-based anonymizers is that this mismatch was mitigated by the OHNN-based anonymizers trained using \textit{VoxCeleb 2}, which contains large-scale, multi-speaker, and multi-language data.

\section{Conclusions}
\label{sec:conclusion}
This paper has proposed a novel OHNN-based speaker anonymization approach that rotates original speaker vectors into anonymized ones with a distribution following the original speaker vector space. 
Towards good privacy protection and voice distinctiveness, AAM/w-AAM and cosine similarity loss functions were introduced to encourage the generation of distinctive anonymized speaker vectors.
Experiments on English VPC datasets demonstrated that the proposed model protects speaker privacy while maintaining speech content: it achieved competitive performance under all attack scenarios in terms of privacy and utility metrics.
Comparison of the cosine similarities between pairs of speaker vectors extracted from the generated speech with a commonly used selection-based anonymizer and the OHNN-based anonymizer further verified that our proposed method can effectively reduce privacy leakage when facing different attackers, while improving the diversity of anonymized speakers.
Experiments on the Mandarin \textit{AISHELL-3} datasets demonstrated that our OHNN-based anonymizer is more robust to the language mismatch scenario than the selection-based methods and can be adopted for this unseen-language anonymization task directly.

To further improve the privacy protection performance under various attack scenarios, our future work will investigate the training loss. One potential direction is to optimize the distance between the original and anonymized speaker vectors by integrating a proxy ASV evaluation model into the training process i.e., using an ASV to measure $\mathcal{L}_{s}$ in Eq.~(\ref{eqn:general_objective}) on original and anonymized speech waveforms. Such a training scheme is closer to how attackers infringe on the speaker's identity. 
{Additionally, we are considering extending the OHNN-based anonymizer to protect other personal attributes such as age, gender, emotion, and dialect. We previously proposed a system for concealing the gender of a speaker \cite{noe2023hiding}, and we feel the framework can be extended to other attributes as well. Our goal is to achieve controllable voice privacy protection that enables users to customize and control the anonymization process according to their specific privacy needs. 
\ifCLASSOPTIONcaptionsoff
  \newpage
\fi

\bibliographystyle{IEEEtran}
\bibliography{ref}

\appendices
\section{Detailed results}
\label{sec:app}
\begin{table}[b!]
  \caption{\textit{Ignorant} EER with AAM+cos, W-AAM+cos of ROH and LOH.}
  \centering
     \renewcommand{\tabcolsep}{0.08cm}
  \resizebox{\columnwidth}{!}{
  \begin{tabular}{|c|c|c||c|c|c|c|c|}
  \hline
 \multirow{3}{*}{	\textbf{Dataset}} &  \multirow{3}{*}{\textbf{Gender}} & \multirow{3}{*}{\textbf{Weight}} & \multicolumn{5}{c|}{\textbf{EER,\%}}\\ \cline{4-8}
 & &  &	 \textbf{S-Select} &	 \multicolumn{2}{c|}{ROH}&  \multicolumn{2}{c|}{LOH} \\ \hline 
& & & &	\makecell[l]{AAM \\ +cos} &  \makecell[l]{ {wAAM} \\  +cos}& \makecell[l]{AAM \\ +cos} &  \makecell[l]{ {wAAM} \\  +cos} \\ \hline \hline
\multirow{2}{*}{LibriSpeech-dev}	&	female	& 0.25 	&47.44 &39.77	&	44.74	&	46.73	&	49.72	\\
							         &	male 	& 0.25	&46.72 & 45.81	&	49.22	&	47.05	&	45.34	\\ \hline
\multirow{2}{*}{VCTK-dev (diff.)}	&	female 	& 0.20	&52.11 &41.55	&	45.54	&	44.97	&	45.48	\\
							         &	male 	& 0.20	&48.04 & 44.07	&	49.53	&	42.18	&	42.68	\\  \hline
\multirow{2}{*}{VCTK-dev (com.)}	&	female 	& 0.05	&47.97 &45.93	&	47.67	&	42.44	&	45.64	\\
							         &	male 	& 0.05	&45.30 &49.29	&	54.42	&	43.87	&	45.30	\\  \hline
\multicolumn{3}{|c||}{\cellcolor{gray!7}Weighted average dev} &	\cellcolor{gray!7} 48.23 &	\cellcolor{gray!7} 43.28	 & \cellcolor{gray!7}	47.60	& \cellcolor{gray!7} 45.19 	& \cellcolor{gray!7} 45.94 \\ \hline\hline
\multirow{2}{*}{LibriSpeech-test}	&	female 	& 0.25	& 41.24 	& 35.77	&	40.51	&	37.77	&	41.61	\\
							         &	male 	& 0.25	& 42.54      & 48.78	&	51.22	&	43.43	&	39.42	\\  \hline
\multirow{2}{*}{VCTK-test (diff.)}	&	female 	& 0.20	& 50.31     & 48.46	&	51.08	&	43.42	&	41.87	\\
							         &	male 	& 0.20	& 48.11      & 47.47	&	52.30	&	45.01	&	43.17	\\  \hline
\multirow{2}{*}{VCTK-test (com.)}	&	female	& 0.05 	& 47.40      & 45.38	&	48.84	&	41.91	&	45.66	\\
							         &	male 	& 0.05	& 47.46      & 50.00	&	55.65	&	41.81	&	41.81	\\ \hline
\multicolumn{3}{|c||}{\cellcolor{gray!7}Weighted average test} &	\cellcolor{gray!7}	45.37 &	\cellcolor{gray!7}	45.09  & \cellcolor{gray!7}	49.83	& \cellcolor{gray!7} 42.17	& \cellcolor{gray!7} 41.63 \\
\hline
  \end{tabular}  }
\end{table}

\begin{table}[b!]
  \caption{\textit{Lazy-informed} EER with AAM+cos, w-AAM+cos of ROH and LOH.}
  \centering
     \renewcommand{\tabcolsep}{0.08cm}
  \resizebox{\columnwidth}{!}{
  \begin{tabular}{|c|c|c||c|c|c|c|c|}
  \hline
 \multirow{3}{*}{	\textbf{Dataset}} &  \multirow{3}{*}{\textbf{Gender}} & \multirow{3}{*}{\textbf{Weight}} & \multicolumn{5}{c|}{\textbf{EER,\%}}\\ \cline{4-8}
 & &  &	 \textbf{S-Select} &	 \multicolumn{2}{c|}{ROH}&  \multicolumn{2}{c|}{LOH} \\ \hline 
& & & &	\makecell[l]{AAM \\ +cos} &  \makecell[l]{ {wAAM} \\  +cos}& \makecell[l]{AAM \\ +cos} &  \makecell[l]{ {wAAM} \\  +cos} \\ \hline \hline
\multirow{2}{*}{LibriSpeech-dev}	&	female	& 0.25  &  29.55 &   42.19	&	43.18	&	51.42	&	57.24	\\
	                                &	male 	& 0.25	&  34.78 &	42.70	&	43.94	&	44.88	&	62.89	\\ \hline
\multirow{2}{*}{VCTK-dev (diff.)}	&	female 	& 0.20	&  24.09 &	29.65	&	32.57	&	38.12	&	58.28	\\
	                                &	male 	& 0.20	&  29.48 &	45.01	&	46.45	&	52.51	&	57.37	\\  \hline
\multirow{2}{*}{VCTK-dev (com.)}	&	female 	& 0.05	&  20.35 &	35.17	&	37.50	&	47.38	&	57.56	\\
	                                &	male 	& 0.05	&  29.63 &	45.87	&	44.73	&	58.40	&	66.53	\\  \hline
\multicolumn{3}{|c||}{\cellcolor{gray!7}Weighted average dev} &	\cellcolor{gray!7} 29.29	& \cellcolor{gray!7} 40.20	 & \cellcolor{gray!7}	41.69	& \cellcolor{gray!7} 47.49 	& \cellcolor{gray!7} 59.31 \\ \hline\hline
\multirow{2}{*}{LibriSpeech-test}	&	female 	& 0.25	& 29.74 &	40.88	&	40.51	&	53.10	&	64.78	\\
	                                &	male 	& 0.25	& 33.18 &	47.88	&	43.65	&	54.34	&	67.04	\\  \hline
\multirow{2}{*}{VCTK-test (diff.)}	&	female 	& 0.20	& 29.32 &	42.39	&	42.70	&	45.11	&	54.53	\\
	                                &	male 	& 0.20	& 30.71 &	59.36	&	54.31	&	44.14	&	61.48	\\  \hline
\multirow{2}{*}{VCTK-test (com.)}	&	female  & 0.05 & 27.75 &	41.91	&	42.49	&	49.71	&	57.51	\\
	                                &	male 	& 0.05	& 29.38 &	54.80	&	51.98	&	48.59	&	62.71	\\ \hline
\multicolumn{3}{|c||}{\cellcolor{gray!7}Weighted average test} & \cellcolor{gray!7} 30.59 &	\cellcolor{gray!7}	47.37  & \cellcolor{gray!7}	45.16	& \cellcolor{gray!7} 49.62	& \cellcolor{gray!7} 62.16 \\
\hline
  \end{tabular} } 
\end{table}

\begin{table}[b!]
  \caption{\textit{Semi-informed} EER with AAM+cos, w-AAM+cos of ROH and LOH.}
  \centering
     \renewcommand{\tabcolsep}{0.08cm}
\resizebox{\columnwidth}{!}{
  \begin{tabular}{|c|c|c||c|c|c|c|c|}
  \hline
 \multirow{3}{*}{	\textbf{Dataset}} &  \multirow{3}{*}{\textbf{Gender}} & \multirow{3}{*}{\textbf{Weight}} & \multicolumn{5}{c|}{\textbf{EER,\%}}\\ \cline{4-8}
 & &  &	 \textbf{S-Select} &	 \multicolumn{2}{c|}{ROH}&  \multicolumn{2}{c|}{LOH} \\ \hline 
& & & &	\makecell[l]{AAM \\ +cos} &  \makecell[l]{ {wAAM} \\  +cos}& \makecell[l]{AAM \\ +cos} &  \makecell[l]{ {wAAM} \\  +cos} \\ \hline \hline
\multirow{2}{*}{LibriSpeech-dev}	&	female	& 0.25  &  11.65  &  	44.89	&	45.88	&	51.14	&	60.65	\\
	                                &	male 	& 0.25	&  4.96 &	40.06	&	43.01	&	36.18	&	61.49	\\ \hline
\multirow{2}{*}{VCTK-dev (diff.)}	&	female 	& 0.20	&  7.52 &	30.38	&	33.58	&	35.26	&	57.33	\\
	                                &	male 	& 0.20	&  6.79 &	43.82	&	42.18	&	44.07	&	61.14	\\  \hline
\multirow{2}{*}{VCTK-dev (com.)}	&	female 	& 0.05	& 7.55 &	33.14	&	36.92	&	36.05	&	52.91	\\
	                                &	male 	& 0.05	& 7.12 &	43.59	&	44.16	&	45.87	&	64.96	\\  \hline
\multicolumn{3}{|c||}{\cellcolor{gray!7}Weighted average dev} &	\cellcolor{gray!7} 7.75 &\cellcolor{gray!7} 39.91	 & \cellcolor{gray!7}	41.42	& \cellcolor{gray!7} 41.79 	& \cellcolor{gray!7} 60.12 \\ \hline\hline
\multirow{2}{*}{LibriSpeech-test}	&	female 	& 0.25	& 4.92	&	34.85	&	37.23	&	41.61	&	59.85	\\
	                                &	male 	& 0.25	& 2.89	&	42.34	&	44.10	&	45.66	&	59.69	\\  \hline
\multirow{2}{*}{VCTK-test (diff.)}	&	female 	& 0.20	& 13.48	&	41.77	&	40.74	&	35.44	&	48.77	\\
	                                &	male 	& 0.20	& 7.75	&	51.49	&	45.75	&	38.17	&	60.73	\\  \hline
\multirow{2}{*}{VCTK-test (com.)}	&	female	& 0.05 &  10.40  & 	41.62	&	39.02	&	38.44	&	55.34	\\
	                                &	male 	& 0.05	&	4.52 &	47.74	&	46.05	&	44.07	&	66.38	\\ \hline
\multicolumn{3}{|c||}{\cellcolor{gray!7}Weighted average test} & \cellcolor{gray!7} 6.94	&	\cellcolor{gray!7}	42.41  & \cellcolor{gray!7}	41.88	& \cellcolor{gray!7} 40.66	& \cellcolor{gray!7} 57.87 \\
\hline
  \end{tabular}  }
\end{table}

\clearpage

\begin{IEEEbiography}[{\includegraphics[width=1in,height=1.25in,clip,keepaspectratio]{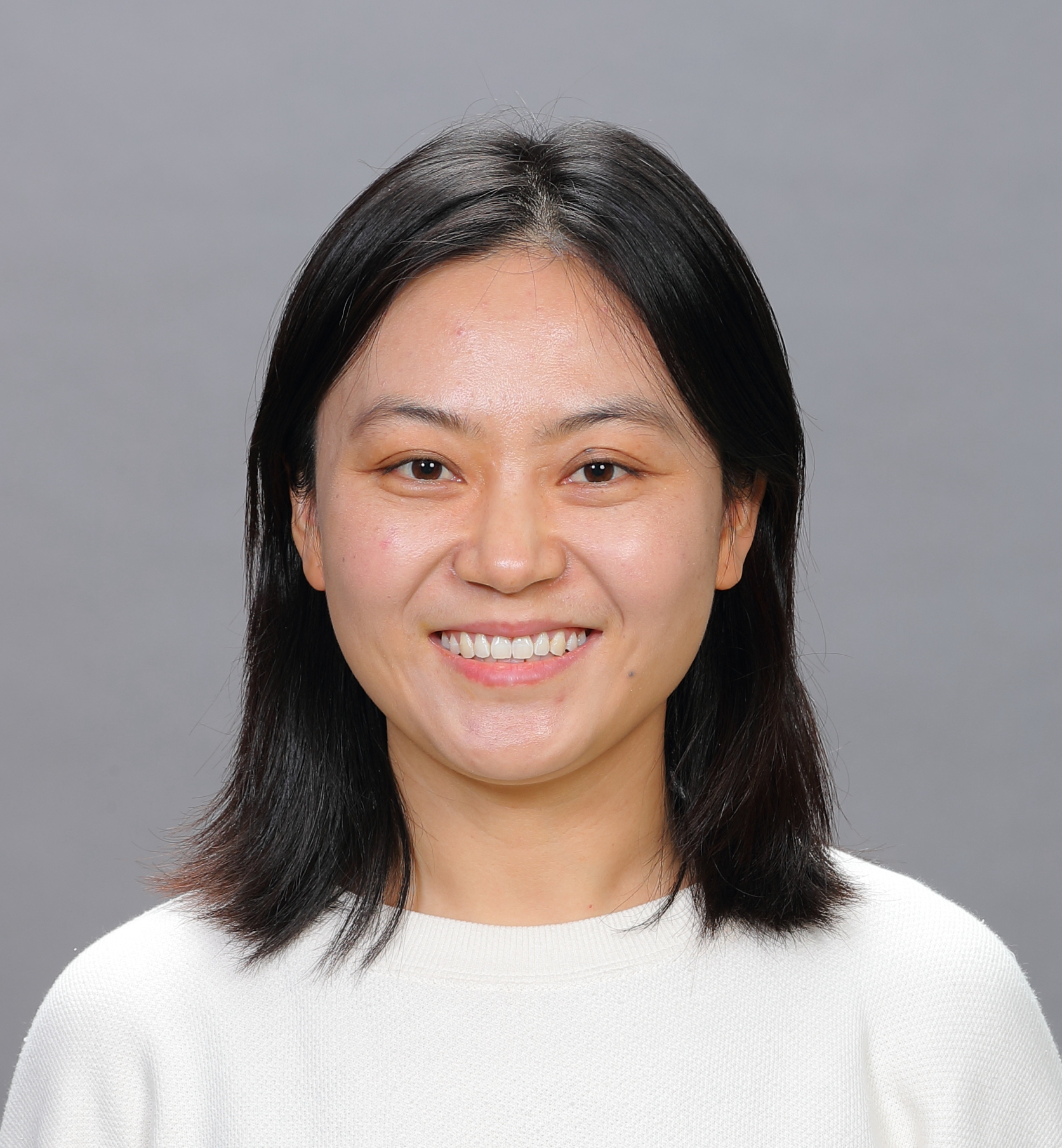}}]{Xiaoxiao Miao} (Member, IEEE) is a project researcher at the National Institute of Informatics (NII), Japan. She received the Ph.D. degree from the Institute of Acoustics, Chinese Academy of Sciences/University Chinese Academy of Sciences, in 2021. Her research interests include speaker and language recognition, speech security, and machine learning. She is a co-organizer of the latest VoicePrivacy challenge. 
\end{IEEEbiography}

\begin{IEEEbiography}[{\includegraphics[width=1in,height=1.25in,clip,keepaspectratio]{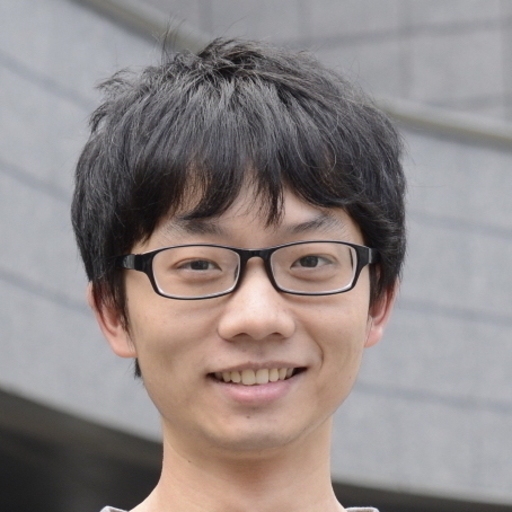}}]{Xin Wang} (Member, IEEE) is a project assistant professor at the National Institute of Informatics (NII), Japan. He received the Ph.D. degree from SOKENDAI/NII, Japan, in 2018. Before that, he received M.S. and B.E degrees from the University of Science and Technology of China and University of Electronic Science and Technology of China in 2015 and 2012, respectively. His research interests include statistical speech synthesis, speech security, and machine learning. He is a co-organizer of the latest ASVspoof and VoicePrivacy challenges. 
\end{IEEEbiography}

\begin{IEEEbiography}[{\includegraphics[width=1in,height=1.25in,clip,keepaspectratio]{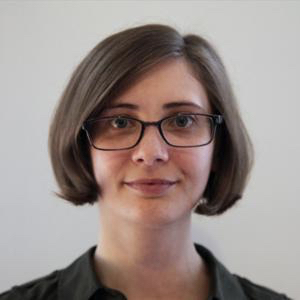}}]{Erica Cooper} (Member, IEEE) received a B.Sc. degree and M.Eng. degree both in electrical engineering and computer science from the Massachusetts Institute of Technology, Cambridge, MA, USA, in 2009 and 2010, respectively. She received a Ph.D. degree in computer science from Columbia University, New York, NY, USA, in 2019. She is currently a Project Assistant Professor with the National Institute of Informatics, Chiyoda, Tokyo, Japan. Her research interests include statistical machine learning and speech synthesis. Dr. Cooper's awards include the 3rd Prize in the CSAW Voice Biometrics and Speech Synthesis Competition, the Computer Science Service Award from Columbia University, and the Best Poster Award in the Speech Processing Courses in Crete.
\end{IEEEbiography}

\begin{IEEEbiography}[{\includegraphics[width=1in,height=1.25in,clip,keepaspectratio]{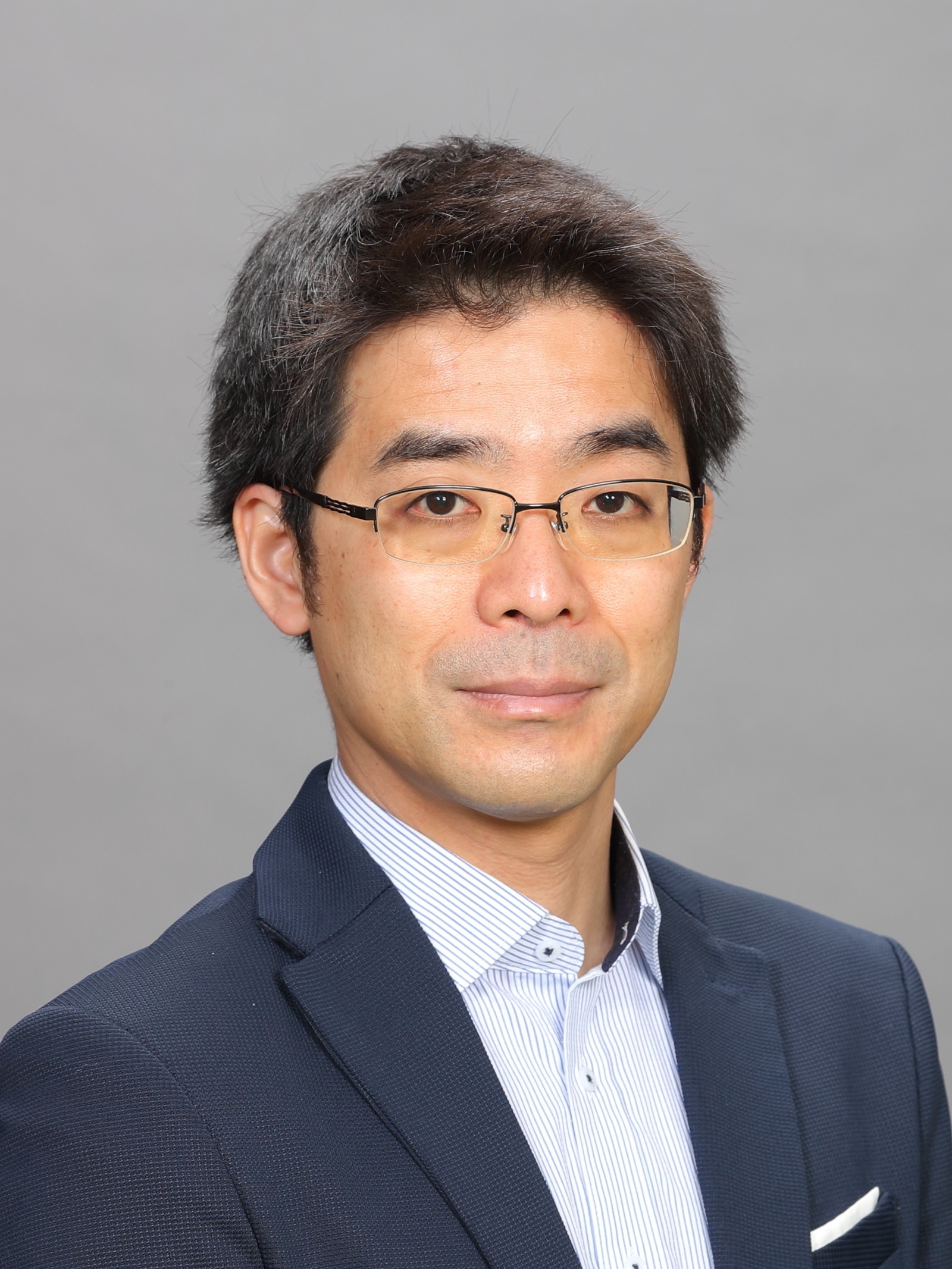}}]
{Junichi Yamagishi} (Senior Member, IEEE) received a Ph.D.\ degree from the Tokyo Institute of Technology (Tokyo Tech), Tokyo, Japan, in 2006. From 2007-2013, he was a research fellow in the Centre for Speech Technology Research (CSTR) at the University of Edinburgh, UK. He was appointed Associate Professor at the National Institute of Informatics, Japan, in 2013. He is currently a Professor at NII, Japan. His research topics include speech processing, machine learning, signal processing, biometrics, digital media cloning, and media forensics. 
He served previously as co-organizer for the bi-annual ASVspoof Challenge and the bi-annual Voice Conversion Challenge. He also served as a member of the IEEE Speech and Language Technical Committee (2013-2019), an Associate Editor of the IEEE/ACM Transactions on Audio Speech and Language Processing (2014-2017),  a chairperson of ISCA SynSIG (2017- 2021), and a Senior Area Editor of the IEEE/ACM TASLP (2019-2023). He is currently a PI of JST-CREST and ANR supported VoicePersonae project.
\end{IEEEbiography}

\begin{IEEEbiography}[{\includegraphics[width=1in,height=1.25in,clip,keepaspectratio]{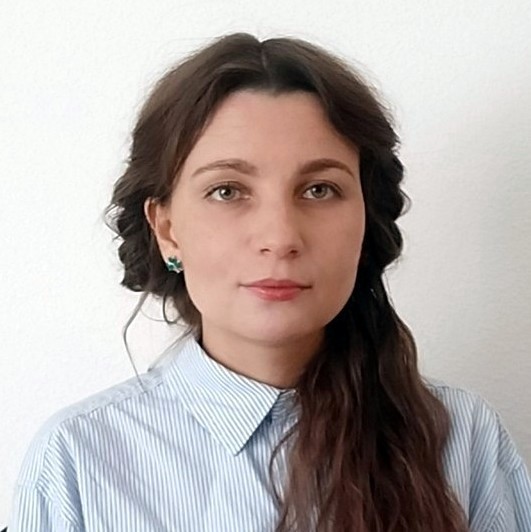}}]{Natalia Tomashenko} (Member, IEEE) is a  researcher at the University of Avignon, France. She received the Ph.D. degree in computer science from the University of Le Mans, France. 
Her  research interests focus on statistical machine learning for speech and language processing with application to automatic speech and speaker recognition, spoken language understanding,  machine translation, and speech privacy. She is an organizer  of the VoicePrivacy challenge. 
\end{IEEEbiography}


\vfill
\clearpage

\newpage
\section{Tables Moved from Main Body to Save Space}

\begin{table}[t]
\centering
\caption{Comparison of the OHNN-based anonymizer and existing methods.}
\label{tab:formulation_comp}
\resizebox{1\linewidth}{!}{
\begin{tabular}{clcccc}
\toprule
 Method & \makecell[c] {Privacy \\ protection} &  \makecell[c]{Speaker \\ diversity}  & \makecell[c]{Naturalness \& \\ Intelligibility}  & \makecell[c]{Automatic\\ mapping}   \\ \midrule
 Selection  \cite{tomashenko2021voiceprivacy,tomashenko2022voiceprivacy}             & \cmark  & \xmark   & \cmark  &  \xmark \\ 
LUT-based \cite{yao22_spsc}        &  \cmark & \xmark   & \cmark  &  \xmark  \\  
 GAN+Selection \cite{meyer2022anonymizing}   &  \cmark & \cmark   & \cmark  &  \xmark  \\  
 Adversarial \cite{chen22_spsc}         & \cmark  & \xmark   & \xmark  &  \cmark \\     
OHNN(ours)            &  \cmark & \cmark   & \cmark & \cmark  \\  

\bottomrule
\end{tabular}
}
\end{table}

\appendices
\section{Detailed results}
\label{sec:app}

\begin{table}[b!]
  \caption{\textit{Ignorant} EER with AAM+cos, W-AAM+cos of ROH and LOH.}
  \centering
     \renewcommand{\tabcolsep}{0.14cm}
  \begin{tabular}{|l|l|c||c|r|r|r|}
  \hline
 \multirow{3}{*}{	\textbf{Dataset}} &  \multirow{3}{*}{\textbf{Gender}} & \multirow{3}{*}{\textbf{Weight}} & \multicolumn{4}{c|}{\textbf{EER,\%}}\\ \cline{4-7}
 & & &	 \multicolumn{2}{c|}{ROH}&  \multicolumn{2}{c|}{LOH} \\ \hline 
& & &	\makecell[l]{AAM \\ +cos} &  \makecell[l]{ {wAAM} \\  +cos}& \makecell[l]{AAM \\ +cos} &  \makecell[l]{ {wAAM} \\  +cos} \\ \hline \hline
\multirow{2}{*}{LibriSpeech-dev}	&	female	& 0.25 	&		39.77	&	44.74	&	46.73	&	49.72	\\
							&	male 	& 0.25	&		45.81	&	49.22	&	47.05	&	45.34	\\ \hline
\multirow{2}{*}{VCTK-dev (diff.)}	&	female 	& 0.20	&		41.55	&	45.54	&	44.97	&	45.48	\\
							&	male 	& 0.20	&		44.07	&	49.53	&	42.18	&	42.68	\\  \hline
\multirow{2}{*}{VCTK-dev (com.)}	&	female 	& 0.05	&		45.93	&	47.67	&	42.44	&	45.64	\\
							&	male 	& 0.05	&		49.29	&	54.42	&	43.87	&	45.30	\\  \hline
\multicolumn{3}{|c||}{\cellcolor{gray!7}Weighted average dev} &		\cellcolor{gray!7} 43.28	 & \cellcolor{gray!7}	47.60	& \cellcolor{gray!7} 45.19 	& \cellcolor{gray!7} 45.94 \\ \hline\hline
\multirow{2}{*}{LibriSpeech-test}	&	female 	& 0.25	&		35.77	&	40.51	&	37.77	&	41.61	\\
							&	male 	& 0.25	&		48.78	&	51.22	&	43.43	&	39.42	\\  \hline
\multirow{2}{*}{VCTK-test (diff.)}	&	female 	& 0.20	&		48.46	&	51.08	&	43.42	&	41.87	\\
							&	male 	& 0.20	&		47.47	&	52.30	&	45.01	&	43.17	\\  \hline
\multirow{2}{*}{VCTK-test (com.)}	&	female	& 0.05 	&		45.38	&	48.84	&	41.91	&	45.66	\\
							&	male 	& 0.05	&		50.00	&	55.65	&	41.81	&	41.81	\\ \hline
\multicolumn{3}{|c||}{\cellcolor{gray!7}Weighted average test} &		\cellcolor{gray!7}	45.09  & \cellcolor{gray!7}	49.83	& \cellcolor{gray!7} 42.17	& \cellcolor{gray!7} 41.63 \\
\hline
  \end{tabular}  
\end{table}

\begin{table}[b!]
  \caption{\textit{Lazy-informed} EER with AAM+cos, w-AAM+cos of ROH and LOH.}
  \centering
     \renewcommand{\tabcolsep}{0.14cm}
  \begin{tabular}{|l|l|c||c|r|r|r|}
  \hline
 \multirow{3}{*}{	\textbf{Dataset}} &  \multirow{3}{*}{\textbf{Gender}} & \multirow{3}{*}{\textbf{Weight}} & \multicolumn{4}{c|}{\textbf{EER,\%}}\\ \cline{4-7}
 & & &	 \multicolumn{2}{c|}{ROH}&  \multicolumn{2}{c|}{LOH} \\ \hline 
& & &	\makecell[l]{AAM \\ +cos} &  \makecell[l]{ {wAAM} \\  +cos}& \makecell[l]{AAM \\ +cos} &  \makecell[l]{ {wAAM} \\  +cos} \\ \hline \hline
\multirow{2}{*}{LibriSpeech-dev}	&	female	& 0.25       &      	42.19	&	43.18	&	51.42	&	57.24	\\
	                                                  &	male 	& 0.25	&		42.70	&	43.94	&	44.88	&	62.89	\\ \hline
\multirow{2}{*}{VCTK-dev (diff.)}	&	female 	& 0.20	&		29.65	&	32.57	&	38.12	&	58.28	\\
	                                                 	&	male 	& 0.20	&		45.01	&	46.45	&	52.51	&	57.37	\\  \hline
\multirow{2}{*}{VCTK-dev (com.)}	&	female 	& 0.05	&		35.17	&	37.50	&	47.38	&	57.56	\\
	                                                 	&	male 	& 0.05	&		45.87	&	44.73	&	58.40	&	66.53	\\  \hline
\multicolumn{3}{|c||}{\cellcolor{gray!7}Weighted average dev} &		\cellcolor{gray!7} 40.20	 & \cellcolor{gray!7}	41.69	& \cellcolor{gray!7} 47.49 	& \cellcolor{gray!7} 59.31 \\ \hline\hline
\multirow{2}{*}{LibriSpeech-test}	&	female 	& 0.25	&		40.88	&	40.51	&	53.10	&	64.78	\\
	                                                 	&	male 	& 0.25	&		47.88	&	43.65	&	54.34	&	67.04	\\  \hline
\multirow{2}{*}{VCTK-test (diff.)}	&	female 	& 0.20	&		42.39	&	42.70	&	45.11	&	54.53	\\
	                                                 	&	male 	& 0.20	&		59.36	&	54.31	&	44.14	&	61.48	\\  \hline
\multirow{2}{*}{VCTK-test (com.)}	&	female		& 0.05 &		41.91	&	42.49	&	49.71	&	57.51	\\
	                                                 	&	male 	& 0.05	&		54.80	&	51.98	&	48.59	&	62.71	\\ \hline
\multicolumn{3}{|c||}{\cellcolor{gray!7}Weighted average test} &		\cellcolor{gray!7}	47.37  & \cellcolor{gray!7}	45.16	& \cellcolor{gray!7} 49.62	& \cellcolor{gray!7} 62.16 \\
\hline
  \end{tabular}  
\end{table}

\begin{table}[b!]
  \caption{\textit{Semi-informed} EER with AAM+cos, w-AAM+cos of ROH and LOH.}
  \centering
     \renewcommand{\tabcolsep}{0.14cm}
  \begin{tabular}{|l|l|c||c|r|r|r|}
  \hline
 \multirow{3}{*}{	\textbf{Dataset}} &  \multirow{3}{*}{\textbf{Gender}} & \multirow{3}{*}{\textbf{Weight}} & \multicolumn{4}{c|}{\textbf{EER,\%}}\\ \cline{4-7}
 & & &	 \multicolumn{2}{c|}{ROH}&  \multicolumn{2}{c|}{LOH} \\ \hline 
& & &	\makecell[l]{AAM \\ +cos} &  \makecell[l]{ {wAAM} \\  +cos}& \makecell[l]{AAM \\ +cos} &  \makecell[l]{ {wAAM} \\  +cos} \\ \hline \hline
\multirow{2}{*}{LibriSpeech-dev}	&	female	& 0.25       &      	44.89	&	45.88	&	51.14	&	60.65	\\
	                                                  &	male 	& 0.25	&		40.06	&	43.01	&	36.18	&	61.49	\\ \hline
\multirow{2}{*}{VCTK-dev (diff.)}	&	female 	& 0.20	&		30.38	&	33.58	&	35.26	&	57.33	\\
	                                                 	&	male 	& 0.20	&		43.82	&	42.18	&	44.07	&	61.14	\\  \hline
\multirow{2}{*}{VCTK-dev (com.)}	&	female 	& 0.05	&		33.14	&	36.92	&	36.05	&	52.91	\\
	                                                 	&	male 	& 0.05	&		43.59	&	44.16	&	45.87	&	64.96	\\  \hline
\multicolumn{3}{|c||}{\cellcolor{gray!7}Weighted average dev} &		\cellcolor{gray!7} 39.91	 & \cellcolor{gray!7}	41.42	& \cellcolor{gray!7} 41.79 	& \cellcolor{gray!7} 60.12 \\ \hline\hline
\multirow{2}{*}{LibriSpeech-test}	&	female 	& 0.25	&		34.85	&	37.23	&	41.61	&	59.85	\\
	                                                 	&	male 	& 0.25	&		42.34	&	44.10	&	45.66	&	59.69	\\  \hline
\multirow{2}{*}{VCTK-test (diff.)}	&	female 	& 0.20	&		41.77	&	40.74	&	35.44	&	48.77	\\
	                                                 	&	male 	& 0.20	&		51.49	&	45.75	&	38.17	&	60.73	\\  \hline
\multirow{2}{*}{VCTK-test (com.)}	&	female		& 0.05 &		41.62	&	39.02	&	38.44	&	55.34	\\
	                                                 	&	male 	& 0.05	&		47.74	&	46.05	&	44.07	&	66.38	\\ \hline
\multicolumn{3}{|c||}{\cellcolor{gray!7}Weighted average test} &		\cellcolor{gray!7}	42.41  & \cellcolor{gray!7}	41.88	& \cellcolor{gray!7} 40.66	& \cellcolor{gray!7} 57.87 \\
\hline
  \end{tabular}  
\end{table}

\begin{table*}[b!]
  \caption{\textit{Ignorant} EER,\% achieved by $ASV_\text{eval}$ on data processed by \textbf{B1.a}, \textbf{B1.b}, \textbf{B2} , \textbf{S-Select},   \textbf{S-ROH}, \textbf{S-LOH},vs.\ EER achieved by $ASV_\text{eval}$ on the original (Orig.) unprocessed data.}
  \centering
  \begin{tabular}{|l|l|c||c||r|r|r|r|r|r|}
\Xhline{0.7pt}
 \multirow{2}{*}{	\textbf{Dataset}} &  \multirow{2}{*}{\textbf{Gender}} & \multirow{2}{*}{\textbf{Weight}} & \multicolumn{7}{c|}{\textbf{EER,\%}}\\ \cline{4-10}
& & &{	\textbf{Orig.}} & \textbf{B1.a} & \textbf{B1.b} & \textbf{B2~}& \textbf{S-Select} & \textbf{S-ROH*} & \textbf{S-LOH*}\\ \hline \hline
\multirow{2}{*}{LibriSpeech-dev}	&	female	& 0.25      &           {8.67}	&	50.43&	53.55&	39.77 & 47.44  & 44.74 & 49.72	\\
	                                                  &	male 	& 0.25	&		1.24	&	57.92&    55.90&	39.13 & 46.72  & 49.22 & 45.34\\ \hline
\multirow{2}{*}{VCTK-dev (different)}	&	female 	& 0.20	&		2.86	&	50.08&	52.90&	28.30 & 52.11 & 45.54	& 45.48\\
	                                                  	&	male 	& 0.20	&		1.44	&	54.04&	54.54&	37.27 & 48.04 & 49.53	& 42.68\\  \hline
\multirow{2}{*}{VCTK-dev (common)}	&	female 	& 0.05	&		2.62	&	49.71&	46.51&	41.28 & 47.97 & 47.67       & 45.64 \\
	                                                  	&	male 	& 0.05	&		{1.43}&	54.99&	54.70&	42.17 & 45.30 & 54.42        & 45.30 \\  \hline
\multicolumn{3}{|c||}{\cellcolor{gray!7}Weighted average dev} &		\cellcolor{gray!7} 3.54	 & \cellcolor{gray!7}	53.14 & \cellcolor{gray!7} 53.91	& \cellcolor{gray!7} 37.01 	& \cellcolor{gray!7} 48.23 & \cellcolor{gray!7} 47.60 & \cellcolor{gray!7} 45.94 \\ \hline\hline
\multirow{2}{*}{LibriSpeech-test}	&	female 	& 0.25	&		7.66	&	47.63&	51.64&	38.32 & 41.24 & 40.51   & 41.61	\\
	                                                  	&	male 	& 0.25	&		{1.11}&	  51.89&	52.12&	41.20 & 42.54	& 51.22   & 39.42\\  \hline
\multirow{2}{*}{VCTK-test (different)}	&	female 	& 0.20	&		4.89	&	47.94&	49.64&	46.40 & 50.31 &51.08 &	 41.87 \\
	                                                  	&	male 	& 0.20	&		{2.07}&	53.79&	55.11&	24.68 & 48.11	& 52.30    & 43.17\\  \hline
\multirow{2}{*}{VCTK-test (common)}	&	female		& 0.05 &		2.89	&	48.27&	51.37&	44.80 & 47.40 & 48.84 & 45.66	\\
	                                                  	&	male 	& 0.05	&		{1.13}&	53.11&	53.67&	39.27 & 47.46	 & 55.65    & 41.81 \\ \hline
\multicolumn{3}{|c||}{\cellcolor{gray!7}Weighted average test} &		\cellcolor{gray!7}	3.79  & \cellcolor{gray!7}	50.29	& \cellcolor{gray!7} 52.14	& \cellcolor{gray!7} 38.29  & \cellcolor{gray!7} 45.37  & \cellcolor{gray!7} 49.83 & \cellcolor{gray!7} 41.63 \\
\Xhline{0.7pt}
  \end{tabular}  
\end{table*}

\begin{table*}[b!]
  \caption{\textit{Lazy-informed} EER,\% achieved by $ASV_\text{eval}$ on data processed by \textbf{B1.a}, \textbf{B1.b}, \textbf{B2} ,  \textbf{S-Select},   \textbf{S-ROH}, \textbf{S-LOH},vs.\ EER achieved by $ASV_\text{eval}$ on the original (Orig.) unprocessed data.}
  \centering
  \begin{tabular}{|l|l|c||c||r|r|r|r|r|r|}
\Xhline{0.7pt}
 \multirow{2}{*}{	\textbf{Dataset}} &  \multirow{2}{*}{\textbf{Gender}} & \multirow{2}{*}{\textbf{Weight}} & \multicolumn{7}{c|}{\textbf{EER,\%}}\\ \cline{4-10}
& & &{	\textbf{Orig.}} & \textbf{B1.a} & \textbf{B1.b} & \textbf{B2~}& \textbf{S-Select} & \textbf{S-ROH*} & \textbf{S-LOH*}\\ \hline \hline
\multirow{2}{*}{LibriSpeech-dev}	&	female	& 0.25      &            {8.67}	&	36.22&	31.25&	43.32& 29.55  & 43.18 & 57.24	\\
	                                                  &	male 	& 0.25	&		1.24	&	34.47&	32.61&	44.75& 34.78  & 43.94	& 62.89\\ \hline
\multirow{2}{*}{VCTK-dev (different)}	&	female 	& 0.20	&		2.86	&	26.05&	22.12&	55.76& 24.09 & 32.57	& 58.28\\
	                                                  	&	male 	& 0.20	&		1.44	&	30.97&	24.12&	32.90& 29.48 & 46.45	& 57.37\\  \hline
\multirow{2}{*}{VCTK-dev (common)}	&	female 	& 0.05	&		2.62	&	27.91&	17.15&	40.99& 20.35	& 37.50       & 57.56 \\
	                                                  	&	male 	& 0.05	&		{1.43}&	33.05&	26.50&	40.17& 29.63	& 44.73        & 65.53 \\  \hline
\multicolumn{3}{|c||}{\cellcolor{gray!7}Weighted average dev} &		\cellcolor{gray!7} 3.54	 & \cellcolor{gray!7}	32.12	& \cellcolor{gray!7} 27.39 	& \cellcolor{gray!7} 43.80 & \cellcolor{gray!7} 29.29 & \cellcolor{gray!7} 41.69 & \cellcolor{gray!7} 59.31 \\ \hline\hline
\multirow{2}{*}{LibriSpeech-test}	&	female 	& 0.25	&		7.66	&	32.30&	27.55&	39.78& 29.74 & 40.51   & 64.78	\\
	                                                  	&	male 	& 0.25	&		{1.11}&	 36.08&	34.97&	40.09& 33.18	& 43.65   & 67.04\\  \hline
\multirow{2}{*}{VCTK-test (different)}	&	female 	& 0.20	&		4.89	&	31.84&	22.69&	58.85& 29.32 &42.70 &	 54.53 \\
	                                                  	&	male 	& 0.20	&		{2.07}&	30.88&	25.03&	43.92& 30.71	& 54.31    & 61.48\\  \hline
\multirow{2}{*}{VCTK-test (common)}	&	female		& 0.05 &		2.89	&	31.79&	22.25&	53.18& 27.75 & 42.49 & 57.51	\\
	                                                  	&	male 	& 0.05	&		{1.13}&	31.92&	24.58&	37.29& 29.38	 & 51.98    & 62.71 \\ \hline
\multicolumn{3}{|c||}{\cellcolor{gray!7}Weighted average test} &		\cellcolor{gray!7}	3.79  & \cellcolor{gray!7}	32.82	& \cellcolor{gray!7} 27.51	& \cellcolor{gray!7} 45.04  & \cellcolor{gray!7} 30.59  & \cellcolor{gray!7} 45.16 & \cellcolor{gray!7} 62.16 \\

\Xhline{0.7pt}
  \end{tabular}  
\end{table*}

\begin{table*}[b!]
  \caption{\textit{Semi-informed} EER,\% achieved by $ASV_\text{eval}^\text{anon}$ on data processed by \textbf{B1.a}, \textbf{B1.b}, \textbf{B2} , \textbf{S-Select},   \textbf{S-ROH}, \textbf{S-LOH},vs.\ EER achieved by $ASV_\text{eval}$ on the original (Orig.) unprocessed data.}
  \centering
  \begin{tabular}{|l|l|c||c||r|r|r|r|r|r|}
\Xhline{0.7pt}
 \multirow{2}{*}{	\textbf{Dataset}} &  \multirow{2}{*}{\textbf{Gender}} & \multirow{2}{*}{\textbf{Weight}} & \multicolumn{7}{c|}{\textbf{EER,\%}}\\ \cline{4-10}
& & &{	\textbf{Orig.}} & \textbf{B1.a} & \textbf{B1.b} & \textbf{B2~}& \textbf{S-Select} & \textbf{S-ROH*} & \textbf{S-LOH*}\\ \hline \hline
\multirow{2}{*}{LibriSpeech-dev}	&	female	& 0.25      &            {8.67}	&	17.76&	19.03&	11.36  & 11.65 & 45.88 &	60.65\\
	                                                  &	male 	& 0.25	&		1.24	&	6.37	&	5.59	&	1.40  & 4.96	& 43.01 & 61.49\\ \hline
\multirow{2}{*}{VCTK-dev (different)}	&	female 	& 0.20	&		2.86	&	12.46&	8.25	&	6.68 & 7.52	& 33.58 & 57.33\\
	                                                  	&	male 	& 0.20	&		1.44	&	9.33	&	6.01	&	6.35 & 6.79	& 42.18 & 61.14\\  \hline
\multirow{2}{*}{VCTK-dev (common)}	&	female 	& 0.05	&		2.62	&	13.95&	9.01	&	5.81	& 7.55       & 36.92 & 52.91\\
	                                                  	&	male 	& 0.05	&		{1.43}&	13.11&	9.40	&	8.83	& 7.12        & 44.16 & 64.96\\  \hline
\multicolumn{3}{|c||}{\cellcolor{gray!7}Weighted average dev} &		\cellcolor{gray!7} 3.54	 & \cellcolor{gray!7}	11.74	& \cellcolor{gray!7} 9.93 	& \cellcolor{gray!7} 6.53 & \cellcolor{gray!7} 7.75 & \cellcolor{gray!7} 41.42 & \cellcolor{gray!7} 60.12\\ \hline\hline
\multirow{2}{*}{LibriSpeech-test}	&	female 	& 0.25	&		7.66	&	12.04&	9.49	&	7.12 & 4.92   & 37.23 & 59.85\\
	                                                  	&	male 	& 0.25	&		{1.11}&	  8.91&	7.80	&	1.11	& 2.89   & 44.10 & 59.69 \\  \hline
\multirow{2}{*}{VCTK-test (different)}	&	female 	& 0.20	&		4.89	&	16.00&	10.91&	16.92 &13.48 &	 40.74 & 48.77 \\
	                                                  	&	male 	& 0.20	&		{2.07}&	10.05&	7.52	&	7.69	& 7.75    & 45.75 & 60.73 \\  \hline
\multirow{2}{*}{VCTK-test (common)}	&	female		& 0.05 &		2.89	&	17.34&	15.32&	10.98 & 10.40 & 39.02 & 55.34 \\
	                                                  	&	male 	& 0.05	&		{1.13}&	9.89	&	8.19	&	4.80	 & 4.52    & 46.05 & 66.38 \\ \hline
\multicolumn{3}{|c||}{\cellcolor{gray!7}Weighted average test} &		\cellcolor{gray!7}	3.79  & \cellcolor{gray!7}	11.81	& \cellcolor{gray!7} 9.18	& \cellcolor{gray!7} 7.77  & \cellcolor{gray!7} 6.94  & \cellcolor{gray!7} 41.88 & \cellcolor{gray!7} 57.87 \\
\Xhline{0.7pt}
  \end{tabular}  
\end{table*}

\begin{table}[h!]
  \caption{Gain of voice distinctiveness $G_{\text{VD}}$ achieved on data processed by \textbf{B1.a}, \textbf{B1.b}, or \textbf{B2}, \textbf{S-Select},   \textbf{S-ROH}, \textbf{S-LOH}}\label{tab:second-results}
  \renewcommand{\tabcolsep}{0.11cm}
  \centering
  \begin{tabular}{|l|l|c||r|r|r|r|r|r|r}
\Xhline{0.7pt} 
\multirow{2}{*}{\textbf{Dataset}} & \multirow{2}{*}{\textbf{Gender}}  & \multirow{2}{*}{\textbf{Weight}} & \multicolumn{6}{c|}{ $\boldsymbol{G}_{\text{VD}}$} \\ \cline{4-9}
& & &  \textbf{B1.a} & \textbf{B1.b} & \textbf{B2~}& \textbf{S-Select} & \textbf{S-ROH*} & \textbf{S-LOH*}\\ \hline \hline
\multirow{2}{*}{LibriSpeech-dev}	&	female & 0.25	&	-9.15	&	-4.92	&	-1.94&	-5.80&	-1.70&	-1.76	\\
							&	male  & 0.25	&	-8.94	&	-6.38	&	-1.65&	-7.88&	-0.73&	-1.53\\ \hline
\multirow{2}{*}{VCTK-dev (different)}	&	female	 & 0.20 &	-8.82	&	-5.94	&	-1.32&	-7.56&	-2.64&	-3.95	\\
							&	male  & 0.20	&	-12.61&	-9.38	&	-2.18&	-11.50&	-3.59&	-3.79	\\  \hline
\multirow{2}{*}{VCTK-dev (common)}	&	female  & 0.05	&	-7.56	&	-4.17	&	-1.14	&	-4.98&	-0.61&	-1.88\\
							&	male  & 0.05	&	-10.37&	-6.99	&	-1.32	&	-8.34&	-0.60&	-1.04\\  \hline
\multicolumn{3}{|c||}{\cellcolor{gray!7}Weighted average dev}	& \cellcolor{gray!7}	-9.71 &	\cellcolor{gray!7}	-6.44 &\cellcolor{gray!7}	-1.72& \cellcolor{gray!7}	-7.90 & \cellcolor{gray!7}	-1.92 & \cellcolor{gray!7}	-2.52	\\ \hline\hline
\multirow{2}{*}{LibriSpeech-test}	&	female & 0.25	&	-10.04&	-5.00	&	-1.71&	-5.62&	-0.94&	-1.00	\\
							&	male & 0.25	&	-9.01	&	-6.64	&	-1.74	&	-7.12&	-0.93&	-0.94\\  \hline
\multirow{2}{*}{VCTK-test (different)}	&	female	& 0.20 &	-10.29&	-6.09	&	-1.56&	-9.25&	-3.45&	-4.92	\\
							&	male & 0.20	&	-11.69&	-8.64	&	-1.56&	-10.87&	-2.06&	-3.06\\  \hline
\multirow{2}{*}{VCTK-test (common)}	&	female & 0.05	&	-9.31	&	-5.10	&	-1.59&	-7.39&	-0.94&	-2.49	\\
							&	male & 0.05	&	-10.43&	-6.50	&	-1.36&	-8.24&	-0.42&	-0.84	\\ \hline
\multicolumn{3}{|c||}{\cellcolor{gray!7}Weighted average test}			&\cellcolor{gray!7} -10.15	&	\cellcolor{gray!7} -6.44	&\cellcolor{gray!7}	-1.63	 &\cellcolor{gray!7}	-7.99 &\cellcolor{gray!7}	-1.64 &\cellcolor{gray!7}	-2.25\\
\Xhline{0.7pt}
  \end{tabular}  
\end{table}

\begin{table}[h!]
  \caption{WER, \% achieved by $ASR_\text{eval}$ on the original (Orig.) unprocessed data with AAM+cos, w-AAM+cos of ROH and LOH.}
  \centering
  \renewcommand{\tabcolsep}{0.11cm}
 \begin{tabular}{|l||r||r|r|r|r|}
 \hline
\textbf{Dataset}  &\textbf{Original}  &   \multicolumn{2}{c|}{ROH} &  \multicolumn{2}{c|}{LOH} \\ \hline 
 & & 	\makecell[l]{AAM \\ +cos} &  \makecell[l]{ {wAAM} \\  +cos}& \makecell[l]{AAM \\ +cos} &  \makecell[l]{ {wAAM} \\  +cos} \\ \hline \hline
LibriSpeech-dev	&		3.82	&	4.71	&	4.74	&	4.73  & 4.82	\\ \hline
VCTK-dev	&	                         10.79&	13.91&	13.90&	13.90 & 13.74	\\ \hline
\cellcolor{gray!7}Average	dev &	\cellcolor{gray!7}  7.30  &  \cellcolor{gray!7}	9.31 & 	\cellcolor{gray!7} 	9.32 &  \cellcolor{gray!7} 9.31 & \cellcolor{gray!7} 9.28	\\ \hline \hline
LibriSpeech-test	&		4.15	&	5.08	&	5.12	&	5.06	 & 5.21\\ \hline
VCTK-test	&		               12.82	&	15.33&	15.36&	15.36 & 15.20	\\ \hline
\cellcolor{gray!7}Average test &	\cellcolor{gray!7} 8.48 & \cellcolor{gray!7}		10.20 &	\cellcolor{gray!7} 10.24 & \cellcolor{gray!7} 10.22 & \cellcolor{gray!7} 10.20 \\
\hline
  \end{tabular}  
\end{table}
\normalsize

\begin{table}[h!]
  \caption{WER, \%achieved by $ASR_\text{eval}^{\text{anon}}$ on the anonymized data with AAM+cos, w-AAM+cos of ROH and LOH}
  \centering
  \renewcommand{\tabcolsep}{0.11cm}
 \begin{tabular}{|l||r||r|r|r|r|}
 \hline
\textbf{Dataset}  &\textbf{Original}  &   \multicolumn{2}{c|}{ROH} &  \multicolumn{2}{c|}{LOH} \\ \hline 
 & &	\makecell[l]{AAM \\ +cos} &  \makecell[l]{ {wAAM} \\  +cos}& \makecell[l]{AAM \\ +cos} &  \makecell[l]{ {wAAM} \\  +cos} \\ \hline \hline
LibriSpeech-dev	&		3.82	&	3.78	&	3.87	&	3.89  & 3.88	\\ \hline
VCTK-dev	&	                         10.79&	11.27&	11.48&	11.06 & 11.16	\\ \hline
\cellcolor{gray!7}Average	dev & \cellcolor{gray!7}   7.31 &  \cellcolor{gray!7}	7.52 & 	\cellcolor{gray!7} 	7.67  &  \cellcolor{gray!7}  7.47& \cellcolor{gray!7} 7.52	\\ \hline \hline
LibriSpeech-test	&	4.15		&	4.08	&	4.12	&	4.22	 & 4.28 \\ \hline
VCTK-test	                &	12.82	&	11.75&	11.91&.      11.76    & 11.61	\\ \hline
\cellcolor{gray!7}Average test &	\cellcolor{gray!7} 8.49 & \cellcolor{gray!7}	7.72	 &	\cellcolor{gray!7} 7.84 & \cellcolor{gray!7}  7.99& \cellcolor{gray!7} 7.94 \\
\hline
  \end{tabular}  
\end{table}
\normalsize

\begin{table}[h!]
   \caption{ WER,\% achieved by $ASR_\text{eval}$ on data processed  by \textbf{B1.a}, \textbf{B1.b}, \textbf{B2} ,  \textbf{S-Select},   \textbf{S-ROH}, \textbf{S-LOH} vs.\ WER achieved by $ASR_\text{eval}$ on the original (Orig.) unprocessed data.}  \centering
  \renewcommand{\tabcolsep}{0.11cm}
  \begin{tabular}{|l|l|r|r|r|r|r|r|r|r|}
\Xhline{0.7pt}
 \multirow{2}{*}{\textbf{Dataset}}    & \multicolumn{7}{c|}{\textbf{WER,\%}}\\ \cline{2-8}
 &{	\textbf{Orig.}} & \textbf{B1.a} & \textbf{B1.b} & \textbf{B2}& \textbf{S-Select} & \textbf{S-ROH} & \textbf{S-LOH}\\ \hline \hline
LibriSpeech-dev	&	3.82	         &	6.39	        &	5.91	       &	8.74	 & 4.50 & 4.74 & 4.82 \\ \hline
VCTK-dev	                 &	10.79	&	15.38	&	15.48	&	25.56 & 12.96 & 13.90 & 13.74	\\ \hline
\cellcolor{gray!7}Average	dev &	\cellcolor{gray!7}  7.31  &  \cellcolor{gray!7}	10.88 & 	\cellcolor{gray!7} 	10.69 &  \cellcolor{gray!7} 17.15  & \cellcolor{gray!7} 8.73 & \cellcolor{gray!7} 9.32  & \cellcolor{gray!7} 9.28 \\ \hline \hline
LibriSpeech-test	&	4.15	        &	6.73	        &	6.08	        &	8.90	 & 4.69 & 5.12 & 5.21  \\ \hline
VCTK-test	                 &	12.82	&	15.23	&	15.60	&	28.15 & 14.86 &15.36 & 15.20	\\ \hline
\cellcolor{gray!7}Average test &	\cellcolor{gray!7} 8.49 & \cellcolor{gray!7}		10.98 &	\cellcolor{gray!7} 10.84 & \cellcolor{gray!7} 18.52 & \cellcolor{gray!7} 9.77 & \cellcolor{gray!7} 10.24  & \cellcolor{gray!7} 10.20\\
\Xhline{0.7pt}
  \end{tabular}  
\end{table}
\normalsize

\begin{table}[h!]
  \caption{ WER,\% achieved by $ASR_\text{eval}^\text{anon}$ on data processed  by \textbf{B1.a}, \textbf{B1.b}, \textbf{B2} ,  \textbf{S-Select},   \textbf{S-ROH}, \textbf{S-LOH} vs.\ WER achieved by $ASR_\text{eval}$ on the original (Orig.) unprocessed data.}
  \centering
  \renewcommand{\tabcolsep}{0.11cm}
  \begin{tabular}{|l|l|r|r|r|r|r|r|r|r|}
\Xhline{0.7pt}
 \multirow{2}{*}{\textbf{Dataset}}    & \multicolumn{7}{c|}{\textbf{WER,\%}}\\ \cline{2-8}
 &{	\textbf{Orig.}} & \textbf{B1.a} & \textbf{B1.b} & \textbf{B2}& \textbf{S-Select} & \textbf{S-ROH*} & \textbf{S-LOH*}\\ \hline \hline
LibriSpeech-dev	&	3.82	         &	4.34	        &	4.19	       &	4.32	 & 3.93 & 3.87 & 3.88 \\ \hline
VCTK-dev	                 &	10.79	&	11.54	&	10.98	&	11.76 & 11.56 & 11.48 & 11.16	\\ \hline
\cellcolor{gray!7}Average	dev &	\cellcolor{gray!7}  7.31  &  \cellcolor{gray!7}	7.94 & 	\cellcolor{gray!7} 	7.59 &  \cellcolor{gray!7} 8.04  & \cellcolor{gray!7} 7.74 & \cellcolor{gray!7} 7.67 & \cellcolor{gray!7} 7.52 \\ \hline \hline

LibriSpeech-test	&	4.15	        &	4.75	        &	4.43	        &	4.58	 & 4.39 & 4.12  & 4.28\\ \hline
VCTK-test	                 &	12.82	&	11.82	&	10.69	&	13.48 & 12.50 &11.91 & 11.61	\\ \hline
\cellcolor{gray!7}Average test &	\cellcolor{gray!7} 8.49 & \cellcolor{gray!7}		8.29 &	\cellcolor{gray!7} 7.56 & \cellcolor{gray!7} 9.03 & \cellcolor{gray!7} 8.44 & \cellcolor{gray!7} 7.84 & \cellcolor{gray!7} 7.94\\
\Xhline{0.7pt}
  \end{tabular}  
\end{table}
\normalsize

\end{document}